\documentclass[useAMS,usenatbib]{mn2e}
\usepackage{graphicx}
\usepackage{amssymb}
\usepackage{wasysym}
\usepackage[usenames]{color}

\def\jref@jnl#1{{\rm#1}}

\def\aj{\jref@jnl{AJ}}                   
\def\actaa{\jref@jnl{Acta Astronomica}}  
\def\araa{\jref@jnl{ARA\&A}}             
\def\apj{\jref@jnl{ApJ}}                 
\def\apjl{\jref@jnl{ApJ}}                
\def\apjs{\jref@jnl{ApJS}}               
\def\ao{\jref@jnl{Appl.~Opt.}}           
\def\apss{\jref@jnl{Ap\&SS}}             
\def\aap{\jref@jnl{A\&A}}                
\def\aapr{\jref@jnl{A\&A~Rev.}}          
\def\aaps{\jref@jnl{A\&AS}}              
\def\azh{\jref@jnl{AZh}}                 
\def\baas{\jref@jnl{BAAS}}               
\def\jrasc{\jref@jnl{JRASC}}             
\def\memras{\jref@jnl{MmRAS}}            
\def\mnras{\jref@jnl{MNRAS}}             
\def\pra{\jref@jnl{Phys.~Rev.~A}}        
\def\prb{\jref@jnl{Phys.~Rev.~B}}        
\def\prc{\jref@jnl{Phys.~Rev.~C}}        
\def\prd{\jref@jnl{Phys.~Rev.~D}}        
\def\pre{\jref@jnl{Phys.~Rev.~E}}        
\def\prl{\jref@jnl{Phys.~Rev.~Lett.}}    
\def\pasp{\jref@jnl{PASP}}               
\def\pasj{\jref@jnl{PASJ}}               
\def\qjras{\jref@jnl{QJRAS}}             
\def\skytel{\jref@jnl{S\&T}}             
\def\solphys{\jref@jnl{Sol.~Phys.}}      
\def\sovast{\jref@jnl{Soviet~Ast.}}      
\def\ssr{\jref@jnl{Space~Sci.~Rev.}}     
\def\zap{\jref@jnl{ZAp}}                 
\def\nat{\jref@jnl{Nature}}              
\def\iaucirc{\jref@jnl{IAU~Circ.}}       
\def\aplett{\jref@jnl{Astrophys.~Lett.}} 
\def\apspr{\jref@jnl{Astrophys.~Space~Phys.~Res.}}
\def\bain{\jref@jnl{Bull.~Astron.~Inst.~Netherlands}} 
\def\fcp{\jref@jnl{Fund.~Cosmic~Phys.}}  
\def\gca{\jref@jnl{Geochim.~Cosmochim.~Acta}}   
\def\grl{\jref@jnl{Geophys.~Res.~Lett.}} 
\def\jcp{\jref@jnl{J.~Chem.~Phys.}}      
\def\jgr{\jref@jnl{J.~Geophys.~Res.}}    
\def\jqsrt{\jref@jnl{J.~Quant.~Spec.~Radiat.~Transf.}}
\def\memsai{\jref@jnl{Mem.~Soc.~Astron.~Italiana}}
\def\nphysa{\jref@jnl{Nucl.~Phys.~A}}   
\def\physrep{\jref@jnl{Phys.~Rep.}}   
\def\physscr{\jref@jnl{Phys.~Scr}}   
\def\planss{\jref@jnl{Planet.~Space~Sci.}}   
\def\procspie{\jref@jnl{Proc.~SPIE}}   

\title[X-ray variability of Swift J1644$+$57]{Long-term  X-ray variability of Swift J1644$+$57}
\author[C. J. Saxton et al.]{Curtis J. Saxton$^{1}$\thanks{E-mail:
cjs2@mssl.ucl.ac.uk (CJS);
roberto.soria@curtin.edu.au (RS);
kw@mssl.ucl.ac.uk (KW)}, 
Roberto Soria$^{2}$,
Kinwah Wu$^{1}$, and N. Paul M. Kuin$^{1}$\\
$^{1}$Mullard Space Science Laboratory, University College London,
Holmbury St Mary, Surrey RH5 6NT, UK\\
$^{2}$International Centre for Radio Astronomy Research, Curtin University, 
GPO Box U1987, Perth, WA 6845, Australia}

\begin{document}

\date{Accepted 2012 February 11. Received 2012 January 31; in original form 2011 September 01}

\pagerange{\pageref{firstpage}--\pageref{lastpage}} \pubyear{2011}

\maketitle

\label{firstpage}

\begin{abstract}
We studied the X-ray timing and spectral variability of
  the X-ray source Sw~J1644$+$57, 
  a candidate for a tidal disruption event.
We have separated the long-term trend (an initial decline followed by a plateau)
  from the short-term dips in the {\it Swift} light-curve.  
Power spectra and Lomb-Scargle periodograms
  hint at possible periodic modulation.
By using structure function analysis, 
  we have shown that the dips were not random
  but occurred preferentially at time intervals 
  $\approx (2.3, 4.5, 9) \times 10^5$~s and their higher-order multiples.
After the plateau epoch, dipping resumed at $\approx (0.7, 1.4)\times10^6$~s
  and their multiples.
We have also found  that the X-ray spectrum became much softer during each 
of the early dip, 
  while the spectrum outside the dips became mildly harder in its long-term evolution. 
We propose that the jet in the system undergoes precession and nutation, 
  which causes the collimated core of the jet
  briefly to go out of our line of sight. 
The combined effects of precession and nutation
  provide a natural explanation for the peculiar patterns of the dips.   
We interpret the slow hardening of the baseline flux 
  as a transition from an extended,
  optically thin emission region to a compact,
  more opaque emission core at the base of the jet.
\end{abstract}


\section{Introduction}

Swift J164449.3$+$573451 (henceforth, Sw~J1644$+$57)
   was discovered by the {\it Swift} Burst Alert Telescope
   \citep[BAT;][]{gehrls2004,barthelmy2005}
   on 2011 March 28 
   \citep{cum11,bur11}.
The long duration of the X-ray emission 
   and flaring events (still ongoing after 8 months),
   and the spatial coincidence with the nucleus
   of a galaxy at redshift $z=0.354$ 
   \citep[luminosity distance $\approx 5.7 \times 10^{27}$cm:
	][]{lev11a,lev11b,tho11}
   made it clear that it was not a Gamma-Ray-Burst, 
   and was instead associated to some kind of sudden accretion event 
   onto a supermassive black hole (BH)
   \citep{blo11b,bur11,lev11b}. In particular, the precise position 
from Very Large Array and Very Long Baseline Array radio observations 
clinched the identification of the X-ray source with 
the nucleus of the host galaxy \citep{zau11}.
The lack of previous evidence of nuclear activity (AGN)
   from the same source, 
   and the short timescale for the outburst rise
   \citep[a few days:][]{kri11} 
   are difficult to reconcile with changes in the large-scale accretion flow
   through an accretion disc, but this scenario is not completely ruled out
   \citep{bur11}.
The most favoured scenario is a tidal disruption event deep inside 
   the sphere of influence of the BH
   \citep{blo11a,blo11b,bur11,can11,soc11}.
The nature of the star being disrupted 
   (and therefore the radius at which the event occurs
   and the timescale on which the
   stellar material circularised into a disc and accreted)
   are still matters of intense debate.
For example, it was argued that 
   if the disrupted star were a white dwarf,
   the smaller tidal disruption radius and characteristic timescale
   would be more consistent with the observed duration 
   of the flares \citep{kro11}. 

There are no direct measurements of the BH mass ($M_\bullet$).
Indirect order-of-magnitude constraints consist of:
   an upper limit to the BH mass that permits tidal disruption 
   events outside the event horizons \citep{ree88}; 
   a minimum variability timescale $\sim 100$ s that sets an upper limit 
   to the light-crossing timescale \citep{blo11b,bur11};
   and well known empirical relations between $M_\bullet$
   and the host galaxian environment
   \citep[whatever causes those trends, e.g.][]{silk1998,jahnke2011}.
For example, based on the optical luminosity of the host galaxy,
\citet{lev11b} estimate a spheroidal stellar mass $ M_{\rm sph} \sim 10^9$--$10^{10} M_{\odot}$.
Using the log-linear spheroidal mass--BH mass relation of \citet{bennert11} (see also \citealt{mag98,lau07,kor11}), this implies
a likely BH mass $2 \times 10^6 M_{\odot} \la M_{\rm BH} \la 10^7 M_{\odot}$ \citep{lev11b};
adopting instead the bent $M_{\rm sph}$--$M_{\rm BH}$ relation of \citet{graham12},
it allows also a lower range of BH masses, $10^5 M_{\odot} \la M_{\rm BH} \la 10^7 M_{\odot}$.
The latter range is more consistent with the BH mass $\log (M_\bullet/M_{\odot}) = 5.5 \pm 1.1$
estimated by \citet{mil11}, based on empirical ``fundamental plane'' relations
between radio and X-ray luminosities of accreting BHs.

The relatively small BH mass implies a bolometric Eddington luminosity
   $\la 10^{45}$~erg~s$^{-1}$,
   and a $0.3$--$10$~keV luminosity a few times smaller.
The average isotropic luminosity in the $0.3$--$10$~keV band
   was $\approx 10^{47}$~erg~s$^{-1}$ in the first 
   few weeks after the initial outburst; several
 months ($\sim 10^7$~s) later, 
   it was still $\approx$ a few $\times 10^{45}$~erg~s$^{-1}$,
   well above the Eddington limit.
This is strong evidence that the emission 
   is beamed towards us \citep{blo11b,bur11,lev11b},
   and is probably associated with a relativistic jet, like in blazars.
A relativistic jet has also 
   been invoked to explain the radio transient event, interpreted 
   as an external shock in the gas surrounding Sw~J1644$+$57 
   \citep{bow11,zau11,bur11}.
The Lorentz factor of the jet is highly uncertain,
   with estimates varying from $\Gamma \approx 2$--$5$ 
   \citep{zau11} to $\Gamma \approx 10$--$20$ \citep{blo11b,bur11}.
The rarity of the event,
   with one such outburst\footnote{
   	It has been argued that there may be a second object
	discovered by {\it Swift} in this class \citep{cenko2011},
        and another one discovered by {\it XMM-Newton} \citep{lin11}.
	}
   discovered in 7 years of {\it Swift} operations,
   is consistent with the theoretical expectations
   of tidal disruption rates and jet opening angle 
   $\theta \sim 5^{\circ}$,
   corresponding to $\Gamma \approx 1/\theta \approx 10$.

The origin of the photon emission is still unclear. 
There are clearly at least two prominent peaks
   in the spectral energy distribution:
   one in the far IR and one in the hard X-ray band 
   \citep{blo11b,bur11}.
They can be modelled as direct synchrotron emission 
   (single-component model) from radio to X-rays, 
   with strong dust extinction in the optical/UV band.
Alternatively, the radio/IR peak is the direct synchrotron emission 
   and the X-ray peak is due to inverse Compton scattering of external photons,
   most likely disc photons (two-component blazar model). 
A third possibility is that the X-ray emission is due to
   inverse Compton emission at the base of the jet,
   while the radio/IR synchrotron emission 
   comes from the forward shock at the interface between the head of the jet 
   and the interstellar medium \citep{blo11b}.
 
The X-ray spectrum is well fitted by a simple absorbed power-law, 
   although more complex models were also discussed \citep{bur11}. 
The average photon index during the first $\approx 2\times 10^5$~s 
   is $\Gamma \approx 1.8$ \citep{lev11b,bur11}. 
However, the physical meaning of this average value has to be interpreted 
   more carefully,
   as there are strong variations in hardness and photon index between flares,
   with $\Gamma$ changing between $\approx 1.3$ and $3$ \citep{lev11b}.
In particular, the photon index is harder when the source is brighter
   \citep{ken11,blo11a,lev11b,bur11}.

In the {\it Swift} X-Ray Telescope
\citep[XRT; ][]{burrows2005}
   band, $0.3$--$10$~keV, 
   the decline in flux was initially consistent with the $t^{-5/3}$ decay 
   \citep{lev11b,blo11b}
   expected for the spreading and accretion of 
   fallback material after a tidal disruption event. 
\citet{blo11b} used the early {\it Swift}/XRT observations 
   (up to $t \approx 4.5 \times 10^6$~s)
   to study the power spectral density 
   over the $0.1$--$100$ mHz frequency range: 
   they found no significant feature 
   that is not associated with the orbital period of the spacecraft.
By contrast, a series of {\it XMM-Newton} Target-of-Opportunity observations 
during the initial decline phase suggested the presence of a 4.7-mHz 
($\approx 200$ s) quasi periodic oscillation \citep{mil-stro2011}, 
and showed short-term variability over the $0.4$--$100$ mHz range.  
If the putative quasi-periodic oscillation corresponds to the Keplerian 
frequency at the innermost stable circular orbit of a Schwarzschild 
BH, it would imply a BH mass $\sim 5 \times 10^5 M_{\odot}$ 
\citep{mil-stro2011}.

Possible variability on the $\sim 10^{-7}$--$10^{-4}$ Hz 
frequency range has not been investigated in much detail yet.
The ongoing series of {\it Swift}/XRT monitoring observations, 
which have now followed the source for $\approx 2 \times 10^7$ s, 
offer the best chance for this study.
Luckily, eight months after the outburst, 
   the source is still detected at a flux
   $\approx 10^{-11}$~erg~cm$^{-2}$~s$^{-1}$ 
   in the $0.3$--$10$~keV band, declining only very slowly.
In this paper, we examine the X-ray variability over the $10^4$--$10^7$ s 
timescale,
   looking for phenomenological patterns 
   (characteristic variability timescales, spectral evolution) 
   that can help us test physical scenarios.
In particular, we want to determine whether the dips
   that characterise the X-ray light-curve are random or have some periodicity, 
   and we discuss the X-ray changes during the dips.


\begin{figure}
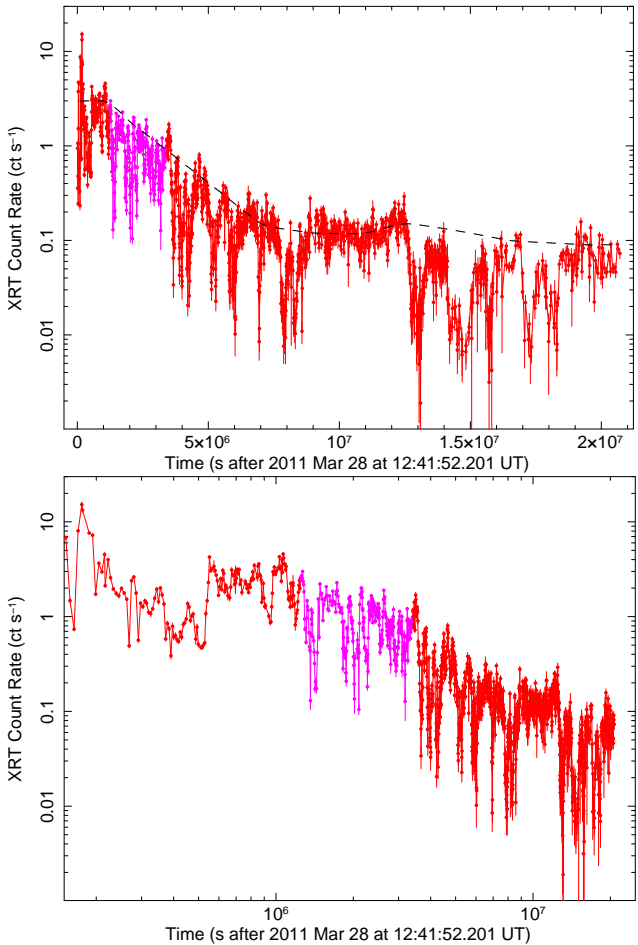

\vspace*{0.15cm}
\begin{center}
\includegraphics[height=84mm,angle=270]{fig1_log.ps}
\includegraphics[height=84mm,angle=270]{fig1_loglog.ps}
\end{center}
\caption{Top panel: {\it Swift}/XRT $0.3$--$10$~keV light-curve
between 2011 March 28 and 2011 November 28, plotted on 
a linear time scale.
Datapoints in red are taken in PC mode; 
those in magenta are in WT mode. 
The dashed curve is a phenomenological function
used to de-trend the data and ``normalise'' the continuum.
Bottom panel: same as in the top panel, but on a logarithmic time scale.}
\label{f1}
\label{fig.lightcurve}
\end{figure}

\section{Data analysis}

Sw~J1644$+$57 has been monitored by the XRT several times a day, every day 
since 2011 March 28.
We used the on-line {\it Swift}/XRT data product generator\footnote{
	Including the new treatment of the vignetting correction,
	introduced after 2011 August 5.}
   \citep{eva07,eva09}
   to extract light-curves in different energy bands, 
   and spectra (including background and ancillary response files) 
   in different time intervals; we selected grade 0--12 events. 
We downloaded the suitable spectral response files for single and double events 
   in photon-counting (PC) mode and window-timing (WT) mode from 
   the latest Calibration Database.  
We used {\small{XSPEC}} Version~12 \citep{arn96} for spectral analysis, 
and standard {\small{FTOOLS}}\footnote{{\tt http://heasarc.gsfc.nasa.gov/ftools}}
   \citep{bla95} tasks for preliminary timing analysis ---
   for example, 
   for defining the time intervals that were used to extract 
   intensity-selected spectra.
For more advanced timing analyses, 
   we used a Lomb-Scargle periodogram and structure function analysis, 
(discussed in Section 3).

A caveat for timing analysis is that the observational data are irregularly 
sampled;
   the duration of each snapshot observation in either WT or PC mode 
   (typically, a few hundred seconds)
   and the temporal gaps between observations are not constant.
Irregular, gappy data sequences are accommodated
   within the formulation of LS-periodograms
   (\S~\ref{s.lomb}),
   where data are at discrete events rather than long temporal bins.
Such imperfections however are unnatural to conventional
   Fourier (\S~\ref{s.fourier})
   and structure function methods (\S~\ref{s.structure}).
There are various strategies that can circumvent this issue: 
  we considered the {\em hyphen method}, the {\em zigzag method} and 
  the {\em trapezoid method}.  
In the hyphen method the gaps are practically omitted. 
The flux levels are set to be locally constant
   for the durations of the data bins. 
In the zigzag method,
   a data point is set at the mid-time of each observational bin,
   and we connect each point linearly to the next consecutive data point.
The numerical light curve is piecewise linear
   and resembles a polygonal landscape.    
In a variant trapezoid method,
   the observational bins are treated as piecewise flat segments 
   as in the hyphen method
   but the gaps are connected directly with diagonal lines. 
   
As the covering factor of the bins is small in the hyphen method, 
   unless the data are numerically bridged, we found that 
   the calculated power spectra and structure functions
   tend to blur any useful information in the noise. 
Instead, we found that the zigzag method tends to provide robust results 
  for power spectra and structure functions. 
The results obtained with the trapezoid method are indistinguishable 
from those of the zigzag method. 
Thus, we present here only the results from analyses done with 
the zigzag method.

\section{Main results}

\subsection{Different phases in the lightcurve}

Figure~\ref{fig.lightcurve}
   shows the XRT $0.3$--$10$~keV light-curve, 
   binned by snapshot (typical exposure duration $\sim$ a few 100 s). 
The luminosity evolution shows 
a series of phases with different phenomenological properties. 
During the first three months after the outburst ---
   in particular, at times $2 \times 10^5~{\rm s} \la t \la 9 \times 10^6$~s ---
   the baseline trend is adequately fitted by an exponential decay,
   with an e-folding timescale of $\approx 2.0 \times 10^6$~s.
Flares and complex dips are superposed onto this trend.
In other work 
   \citep{lev11b},
   a more canonical tidal-disruption decline scaling $\sim t^{-5/3}$ was 
   fitted to the same section of the light-curve;
   the difference is mostly 
   due to a different definition of the baseline. 
Later we will show a change in the variability at $t\approx4.7\times10^6$~s,
   which distinguishes ``early'' and ``late'' epochs
   of the declining stage.
For $t \ga 9 \times 10^6$~s, the decay stopped and the source 
   appeared to settle on a ``plateau''
   (XRT count rate $\approx 0.1$ ct s$^{-1}$), 
   without any dips.
Broader dips resumed for $t \ga 12.5 \times 10^6$~s
   (the ``recent'' epoch),
   but the baseline level has not significantly declined below
   $\approx 0.1$ ct s$^{-1}$.

Because of the different behaviour of the light-curve at different epochs, 
we shall investigate the short-term variability separately in the different 
phases.
We also note that the state transitions
   are each a continuous, gradual evolution of flux, timing and spectral 
   properties over time.
Therefore the choice of start and end times
   defining the different epoch sub-intervals
   are somewhat arbitrary (to within a few $10^5$~s)
   and do not affect the main conclusions of our study.


\begin{figure}
\vspace*{0.15cm}
\begin{center}
	\includegraphics[height=8.4cm,angle=270]{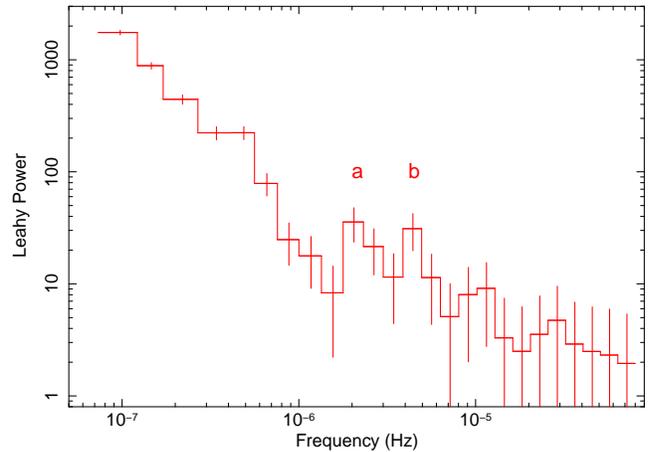}
\end{center}
\caption{
Power spectrum of the {\it Swift}/XRT light curve (combining data 
in WT and PC mode), excluding the first $2 \times 10^5$~s. 
The input data have been binned to 5000 s intervals. 
The power spectrum is normalised \citep{lea83} such that 
its integral gives the squared rms fractional variability
(i.e., the Y axis is in units of (rms)$^2$ Hz$^{-1}$), and the expected 
white noise level $\approx 2$.
Two features at $\mu$Hz frequencies (labelled in the plot) 
are suggestive of periodicities on timescales 
$\approx (2.3 \pm 0.3) \times 10^5$ s and about twice that value.
}
\label{fig.fourier}
\end{figure}

\begin{figure*}
\begin{center}
\begin{tabular}{cc}
 \includegraphics[width=84mm]{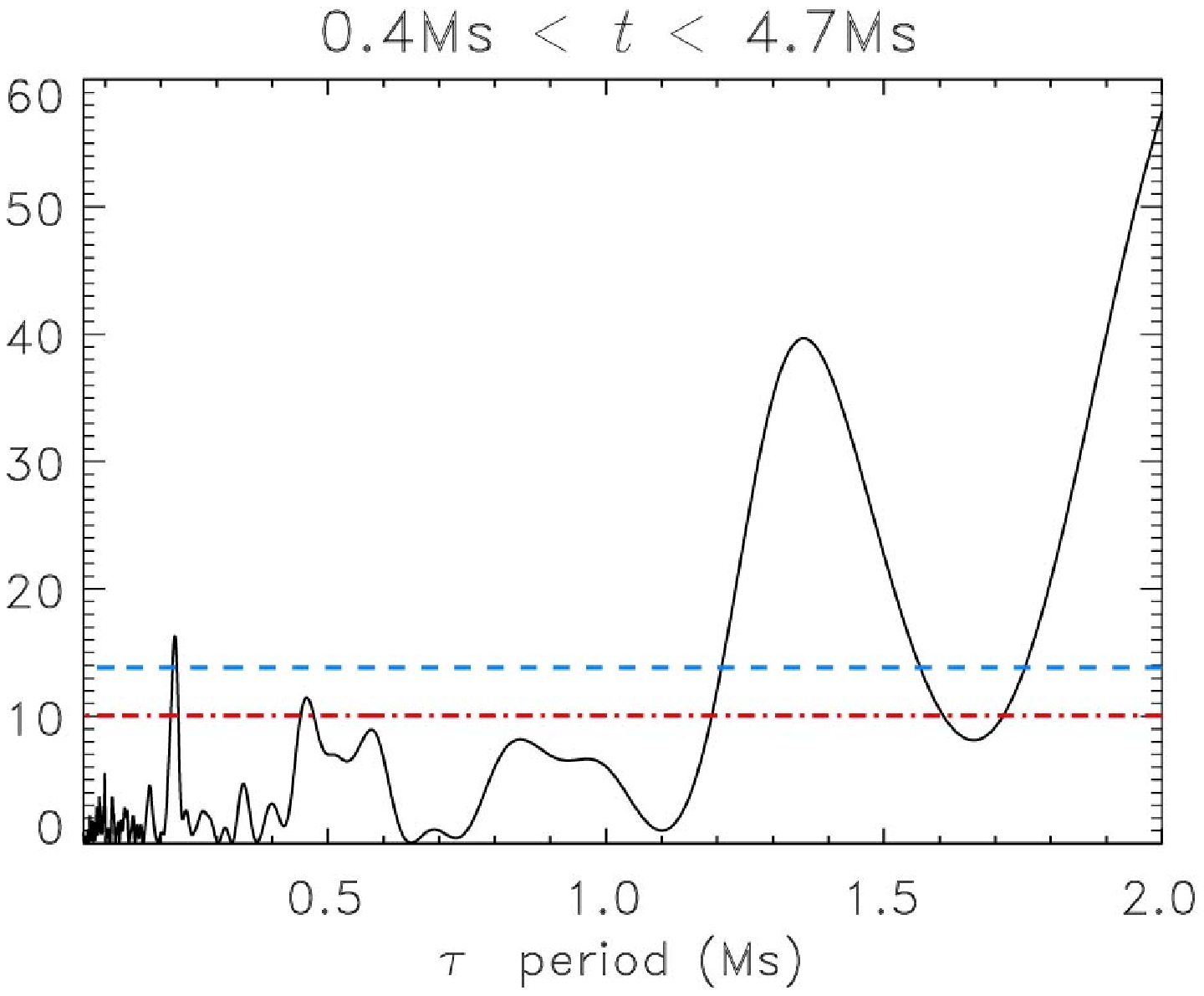}
&\includegraphics[width=84mm]{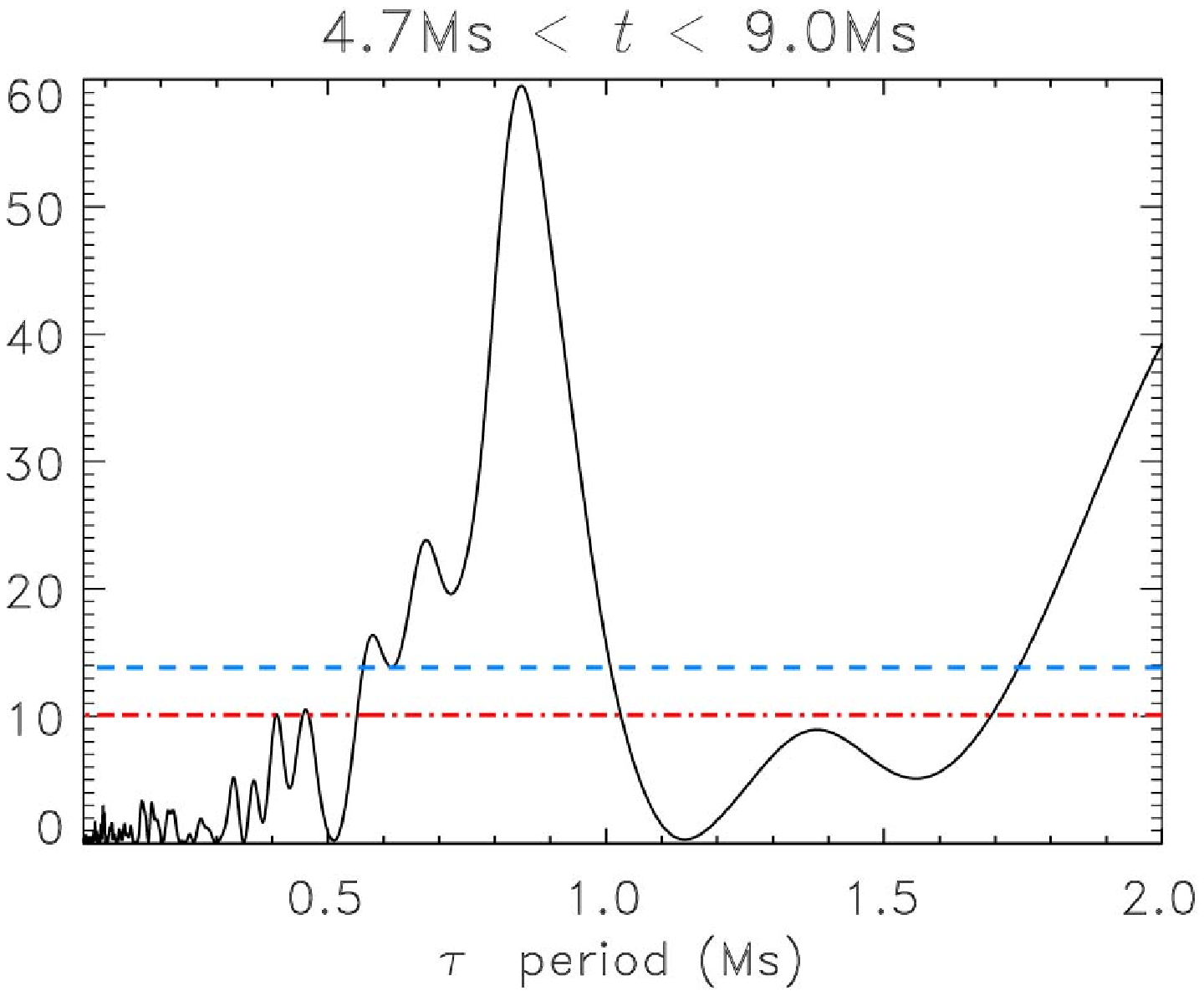}\\
 \includegraphics[width=84mm]{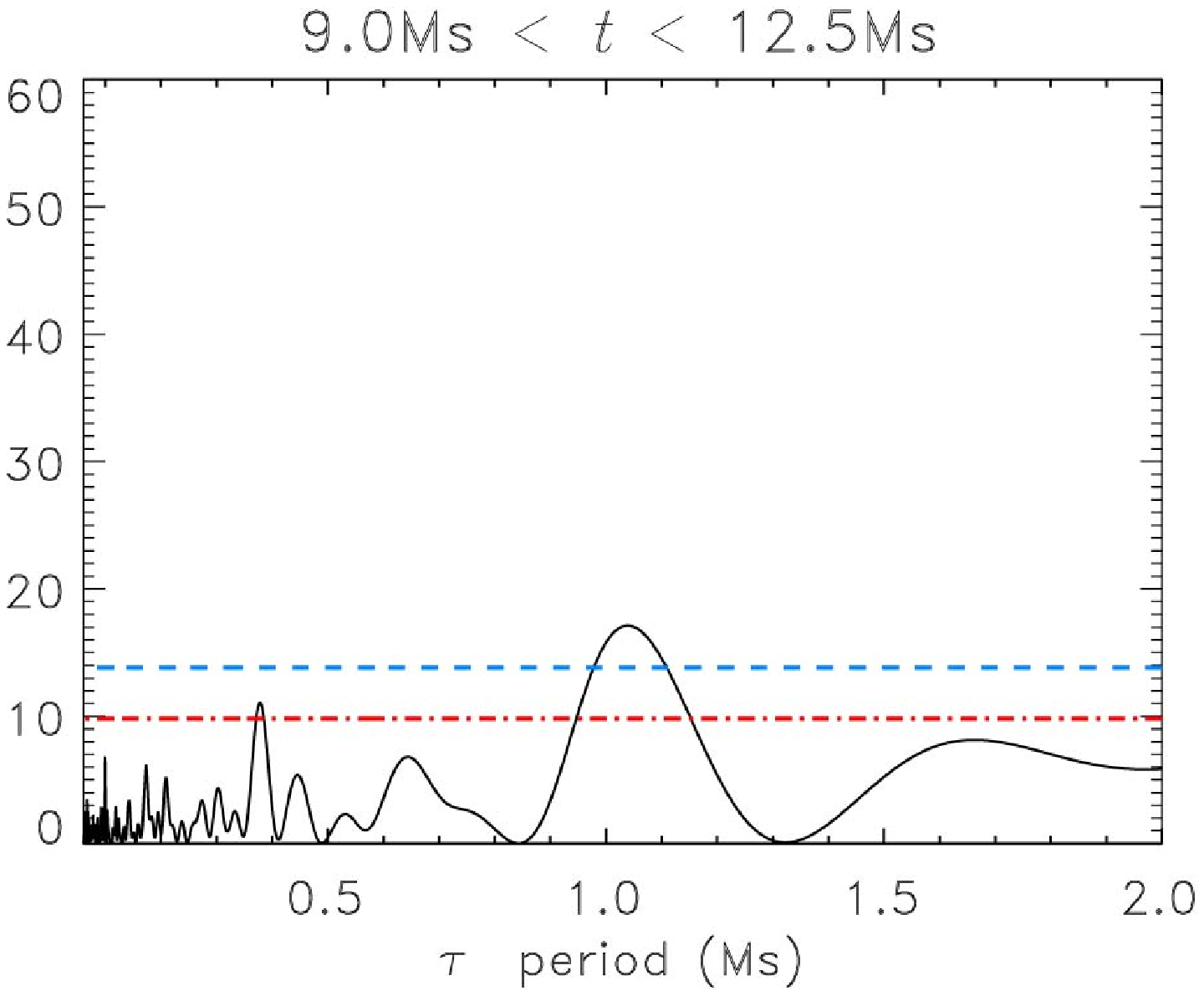}
&\includegraphics[width=84mm]{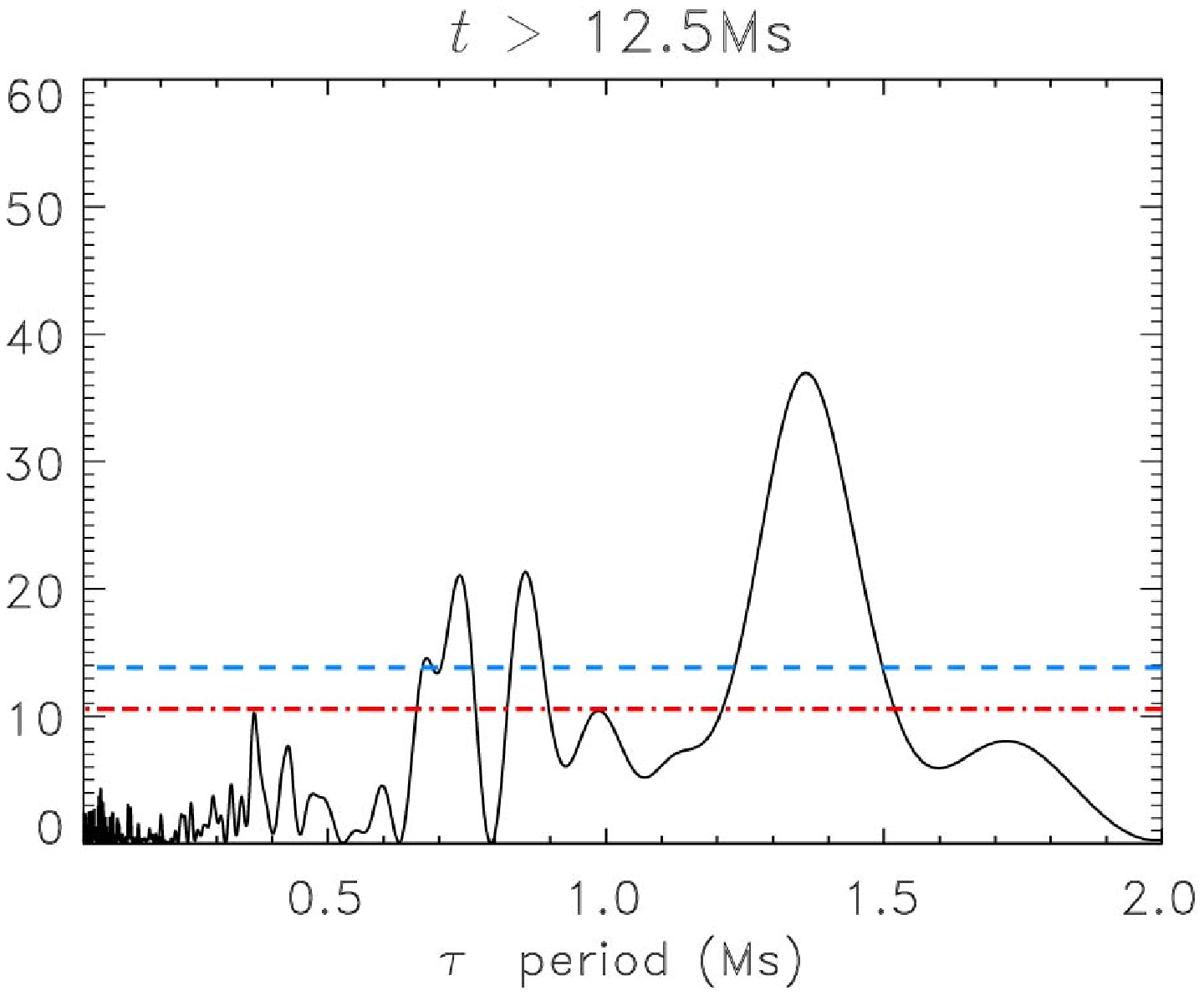}\\
\end{tabular}
\end{center}
\caption{Lomb-Scargle periodograms
   of the light-curve in four time intervals of interest.
The data have not been de-trended.
In the first, second and fourth of these intervals,
   there are plausible periodicities:
   significant at the 1\% FAP level
   (with two definitions of this threshold marked as horizontal lines).
}
\label{fig.lomb}
\label{fig.scargle}
\end{figure*}

\subsection{Search for periodicities}

\subsubsection{Fourier analysis}
\label{s.fourier}

Fourier techniques, as implemented for example in the 
{\small XRONOS} {\it powspec} timing analysis task \citep{ste92},
are widely used for the timing analysis of X-ray lightcurves 
in AGN and X-ray binaries. The standard technique of dividing 
the lightcurve into multiple intervals, calculating 
the power density spectrum in each interval, and taking the mean  
of the power density values (and corresponding standard deviations) 
is applicable only if the process is ergodic \citep{pri88,gui11}, 
so that time averages can be substituted for ensemble averages.
In the case of highly non-stationary events like Sw~J1644$+$57, 
Gamma-ray bursts, or X-ray flares, it does not make sense to subdivide 
the light-curve into sub-intervals, and the power density spectrum 
has to be calculated over the entire duration of the observation 
\citep{gui11}. 
We did that, and then calculated the standard deviation for the power 
density at each frequency bin with the procedure outlined 
in \citet{gui11}.

Figure~\ref{fig.fourier}
   shows the power spectrum of the $0.3$--$10$~keV light-curve,  
   excluding the first $2 \times 10^5$~s
   (that is, after the huge initial flares have subsided).
The root-mean-square (rms) power rises at lower frequencies, 
   as $\sim \nu^{-1}$.
This is mostly due to the long-term dimming trend.
There appear to be two unresolved features at frequencies 
   $\approx 2.2\mu$Hz and $4.5\mu$Hz; however, 
   their power is comparable to the statistical uncertainty. 
This prevents us from firmly concluding
   whether there are true underlying periodicities 
   or whether those features were just statistical noise. 
Moreover, Fourier analysis techniques are better suited 
to time series that are equispaced and with no time gaps. 
Neither condition is true in our case. The complex shape 
of the window and sampling functions may introduce 
side lobes (spectral leakage) in the discrete Fourier transform.
In summary, we consider the features observed in the power density 
spectrum as suggestive of possible periodic signals, but not yet statistically 
significant, because of the shortcomings of the Fourier technique.

\subsubsection{Lomb-Scargle periodogram}
\label{s.lomb}

As an improvement over Fourier power spectrum analyses,
\cite{lomb1976} and \cite{scargle1982}
   introduced a periodogram technique
   that proves to be more robust in the detection of periodicities
   in irregularly sampled light curves.
\cite{press1989} accelerated the method,
   with a numerical modification based on fast Fourier transforms.
The Lomb-Scargle periodogram is equivalent to a best-fit analysis
   of the data with a single sinusoidal function.
The phase and amplitude emerge directly for the implied fit.
The technique is observationally applied to diverse subjects:
   exoplanet detections;
   solar eruptions;
   variable and pulsating stars;
   high-energy accreting systems;
   ultracompact binaries
   \citep[e.g.][]{desidera2011,farrell2010,foullon2009,
	hakala2003,nataf2010,ness2011,omiya2011,qian2010,sarty2009,
	uthas2011,xu2011,wen2006}.

Figure~\ref{fig.lomb}
   presents Lomb-Scargle plots of four stages of the light-curve:
   during the early decline ($4\times10^5< t<4.7\times10^6$~s),
   late decline ($4.7\times10^6<t<9\times10^6$~s),
   the plateau ($9\times10^6<t<1.25\times10^7$~s),
   and the recent post-plateau stage ($t>1.25\times10^7$~s).
We used a freely available {\sc IDL} code\footnote{
{\tt http://astro.uni-tuebingen.de/software/idl/aitlib/timing/}
}
   implementing the formulation of \cite{press1989}
   and \cite{horne1986}.
Power and detection thresholds assume a grid of 1000 evenly spaced frequencies.
For the sake of smoother curves,
   the curve is drawn with $10^4$ intermediate frequencies
   (not involved in the calculation of detection thresholds).
In our plots, peak detection thresholds are marked in blue and red.
Higher peaks are statistically significant at the 1\% level
   (``false alert probability'' FAP$ = 0.01$).
The blue (dashed)
   threshold is $z_0$ in equation (18) of section III(c) of
   \cite{scargle1982}.
The red (dot-dashed) line is the threshold obtained from $10^4$
   ``white noise simulations'' 
   conservatively taking the light-curve's total variance.
The data are not de-trended, so the variance is an overestimate.
The uncorrected decay timescale
   ($t_0\approx2\times10^6$~s)
   during the early and later decline ($t<9\times10^6$~s)
   is probably responsible for artefacts
    at the long-period end of the periodograms
   (top panels of Figure~\ref{fig.lomb}).

\begin{figure}
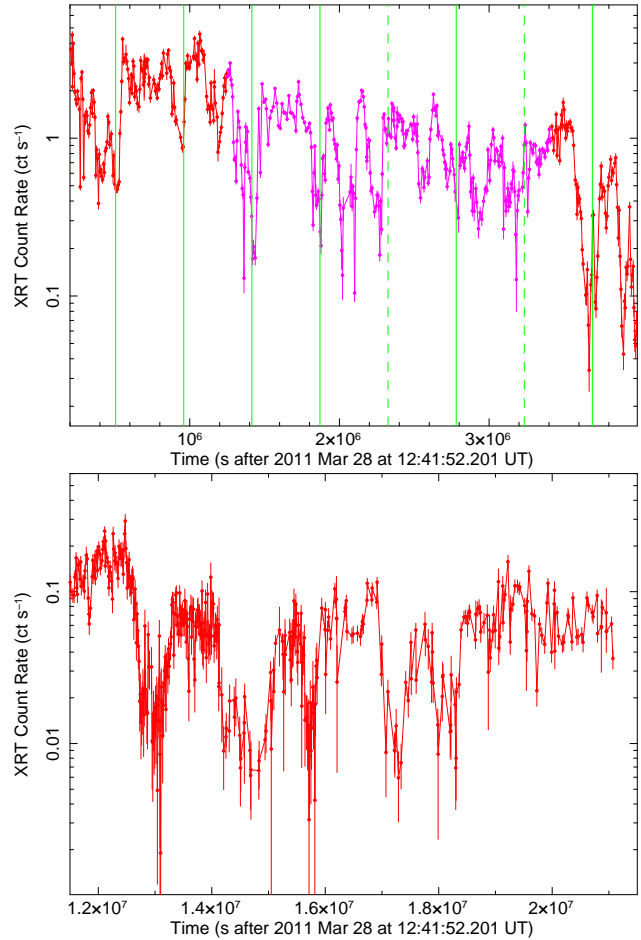

\begin{center}
 \includegraphics[width=62mm,angle=270]{first_dips.ps}\\
 \includegraphics[width=62mm,angle=270]{last_dips.ps}
\end{center}
\caption{Top panel: zoomed-in view of an early section of the light-curve, 
   showing a structure of recurrent dips, many of them 
   consistent with a quasi periodic timescale of $\approx 4.5 \times 10^5$~s 
   (solid green lines).
This kind of dip structure produces the characteristic frequency signal
   detected in the Lomb-Scargle periodograms
   and structure function analyses.
However, some expected dips are skipped or premature
   (dashed green lines).  
Bottom panel:
   zoomed-in view of the last section of the light-curve observed to-date;
   the dips occur on a longer timescale.
}
\label{fig.first_dips}
\label{fig.last_dips}
\end{figure}

At early times ($t<4.7\times10^6$~s)
   there are peaks of varied significance 
   at periods of $\tau\approx0.23$, $0.45$, $0.9$ and $1.4$~Ms.
The first two in this series correspond to the two features tentatively 
identified in the Fourier spectrum (\S~\ref{s.fourier}).
A characteristic spacing $\approx 0.45$Ms between dips is also inferred 
from a visual inspection of the early part 
of the XRT lightcurve (Figure~\ref{fig.first_dips}).
During later weeks of the decline ($4.7\times10^6<t<9\times10^6$~s)
   there is a stronger peak at period $\tau\approx0.9$Ms.
Later, during the plateau stage ($9\times10^6\la t\la12.5\times10^6$~s),
   the power of the peaks is relatively low,
   consistent with the visible steadiness of the light-curve.
There may be a feature at $\tau\approx1.1\times10^6$~s, 
which is near the 1\% significance level,
   and perhaps another, weaker, feature at $\approx0.4\times10^6$~s.
In the very latest times after the plateau, 
   when dipping resumed ($t\ga12.5\times10^6$~s),
   there is a strong peak at $\tau\approx1.4$Ms.
This is consistent with a visual inspection of the last section 
of the lightcurve, as it is the characteristic spacing between major 
quasi-sinusoidal dips (Figure~\ref{fig.last_dips}).

In summary, the Lomb-Scargle analysis gives signs of periodicity
  (which are stronger in some intervals than others),
   and these indications are clearer than from the Fourier analysis.
We note that the Lomb-Scargle periodogram is by construction most sensitive
   to oscillations similar to a sine wave.
Repetitive patterns that are highly non-sinusoidal
   may have their true significance understated.
The dips seen in the early phases 
of the XRT lightcurve (Figure~\ref{fig.first_dips})
   are sharp and brief compared to the non-dip conditions: 
   this is clearly far from a sinusoidal pattern.
This motivates further study with another, broader technique.

\begin{table}
\begin{center}
$
\begin{array}{llll}
\hline \\
\multicolumn{4}{c}{\tau/\tau_0}\\[5pt]
\Delta t_1, 1\sigma
&\Delta t_1, 3\sigma
&\Delta t_2, 1\sigma
&\Delta t_2, 3\sigma
\\[5pt]
\hline
\\
0.514_{-0.038}^{+0.014}	&0.514_{-0.074}^{+0.031} &0.478_{-0.037}^{+0.017} &0.478_{-0.059}^{+0.036}
\\[5pt]
1.037_{-0.014}^{+0.024}	&1.037_{-0.035}^{+0.043} &1.060_{-0.027}^{+0.017} &1.060_{-0.053}^{+0.040}
\\[5pt]
1.555_{-0.026}^{+0.016}	&1.555_{-0.046}^{+0.031} &1.501_{-0.020}^{+0.024} &1.501_{-0.034}^{+0.059}
\\[5pt]
2.034_{-0.036}^{+0.020}	&2.034_{-0.058}^{+0.036} &1.994_{-0.020}^{+0.031} &1.994_{-0.038}^{+0.053}
\\[5pt]
2.455_{-0.024}^{+0.034}	&2.455_{-0.051}^{+0.076} &2.464_{-0.034}^{+0.061} &2.464_{-0.056}^{+0.082}
\\[5pt]
3.052_{-0.052}^{+0.033}	&3.052_{-0.095}^{+0.057} &3.048_{-0.038}^{+0.048} &3.048_{-0.088}^{+0.076}
\\[5pt]
3.558_{-0.027}^{+0.032}	&3.558_{-0.048}^{+0.062} &3.585_{-0.052}^{+0.025} &3.585_{-0.092}^{+0.053}
\\[5pt]
4.002_{-0.032}^{+0.045}	&4.002_{-0.070}^{+0.084} &4.024_{-0.045}^{+0.025} &4.024_{-0.103}^{+0.065}
\\[5pt]
4.498_{-0.044}^{+0.048}	&			 &4.464_{-0.058}^{+0.058} &4.464_{-0.152}^{+0.089}
\\[5pt]
5.066_{-0.190}^{+0.064}	&5.066_{-0.245}^{+0.130} &4.976_{-0.027}^{+0.057} &4.976_{-0.097}^{+0.178}
\\[5pt]
5.536_{-0.138}^{+0.035}	&
\\[5pt]
6.036_{-0.051}^{+0.074}	&			 &6.139_{-0.112}^{+0.068} &6.139_{-0.173}^{+0.122}
\\[5pt]
			&			 &7.629_{-0.139}^{+0.068}
\\[5pt]
8.531_{-0.100}^{+0.108}	&
\\[5pt]
\\
\hline
\end{array}
$
\end{center}
\caption{
Timescale of regular features detected 
in the $S_2$ structure function for data
   extracted in the intervals $\Delta t_1 = [0.4,8.9] \times 10^6$ s
   and $\Delta t_2 = [1.0,8.9] \times 10^6$ s.
Minima are tabulated if their depths exceed $1\sigma$ and $3\sigma$ locally,
   and are not enclosed within the catchment of another local minimum.
To emphasise the periodicity, we express the timescales $\tau$ 
   as multiples of a characteristic $\tau_0=0.445\times10^6$ s.
}
\label{table.dips1}
\label{tab1}
\end{table}

\subsubsection{Structure function technique}
\label{s.structure}

To obtain a more robust test of the presence
  or absence of periodicity in the X-ray luminosity variations, 
  we carried out a more effective alternative analysis based on 
{\it Kolmogorov's structure functions}
   \citep{kol41,kolmogorov1991}.
They are the statistical moments
   of a temporally varying signal,
   with the light curve compared to itself at an offset $\tau$.
The $n^{\rm th}$-order structure function is
\begin{equation}
S_n(\tau) \equiv \langle{
		\left[{	z(t+\tau)-z(t)	}\right]^n
	}\rangle  \ .
\end{equation}
The second order structure function, $S_2$, is particularly useful 
   for the purpose of time series analysis, 
   as it describes the variance of a signal on timescales $\tau$.
In addition to their original applications in the physics of turbulence
   \citep[][and references therein]{brandenburg2011}, 
   structure functions have a venerable history in the diagnosis
   of variability in blazars, other configurations of AGN,
   and various compact systems
   \citep[e.g.][]{simonetti1985,cordes1985,bregman1988,hughes1992,
	brinkmann2000,brinkmann2001,kataoka2001,iyomoto2001,tanihata2003,
	wilhite2008,emmanoulopoulos2010,yusef2011,voevodkin2011}.
In these contexts they are most often used to search for break frequencies
   that are supposed to characterise internal scales of the system,
   though
   \cite{emmanoulopoulos2010}
   warn users to beware of potential artefacts from sparsely sampled data.
(Our data are densely sampled, even during the dips,
   and we are seeking signs of repetition rather than break scales.)
Structure functions
    are also used to diagnose simulations of turbulence in interstellar media
   \citep{boldyrev2002a,kritsuk2007a,kritsuk2009a,schmidt2008},
   and synthetic observables of simulated jets and AGN
   \citep{saxton2002,saxton2010}.
In the Appendix,
   we show the procedures by which we compute
   the structure functions of various orders 
   from data sets with unevenly sampled data.
We also show how the corresponding uncertainties are determined. 
In the rest of this section, we report our main findings.


\begin{figure}
\begin{center}
\begin{tabular}{cc}
\hspace{-0.6cm} \includegraphics[width=9cm]{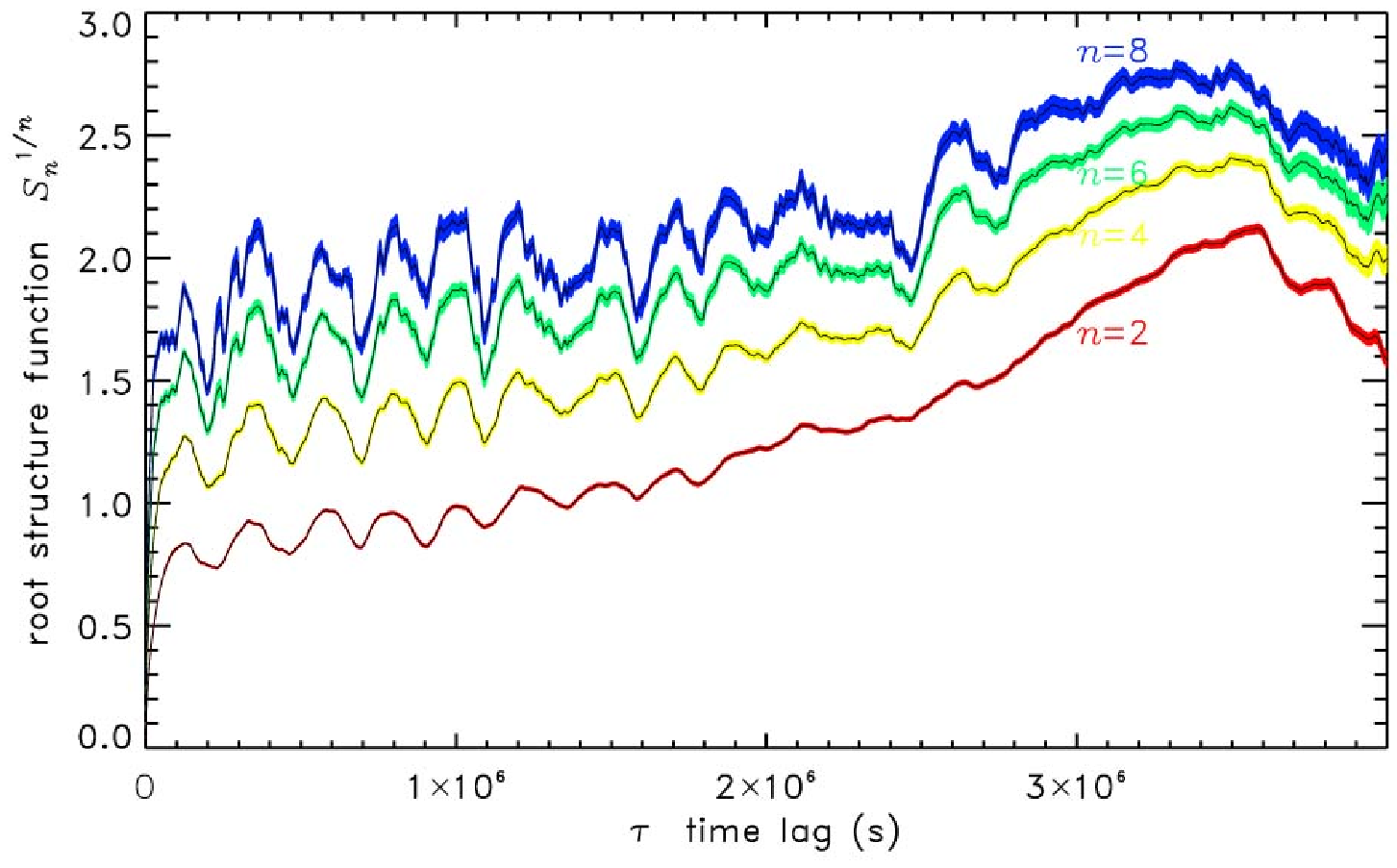}
\\
\hspace{-0.6cm} \includegraphics[width=9cm]{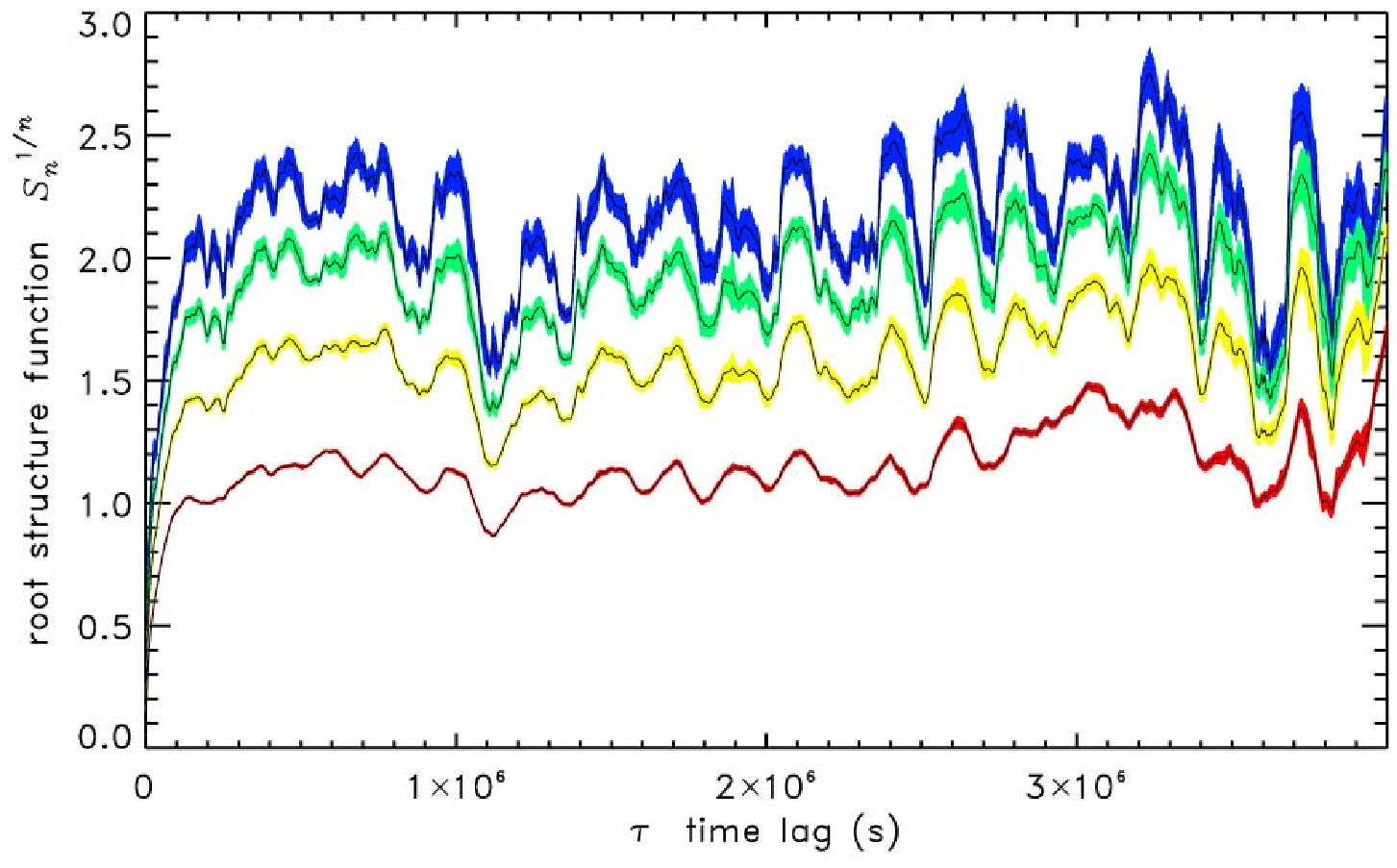}
\\
\hspace{-0.6cm}  \includegraphics[width=9cm]{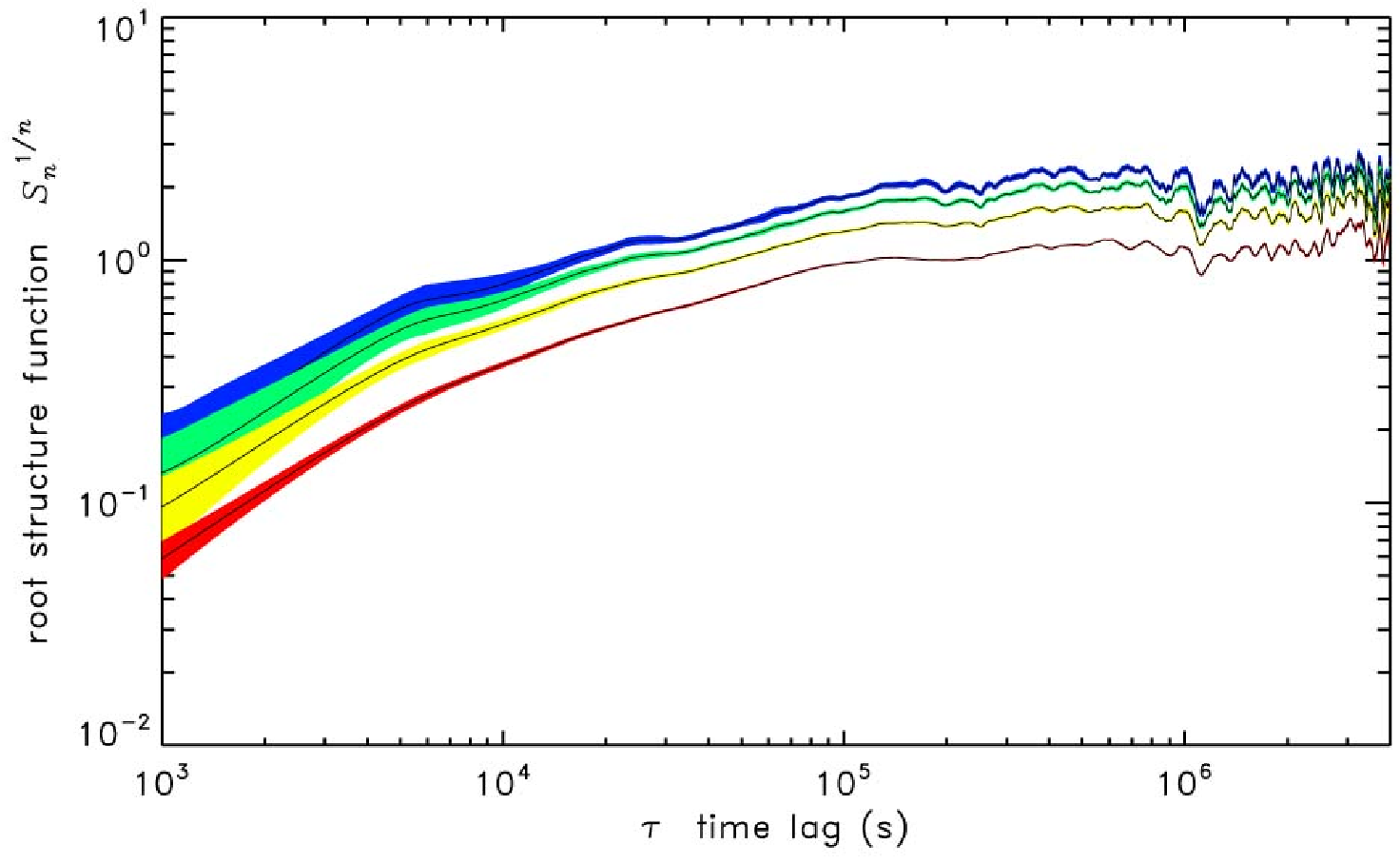}
\end{tabular}
\end{center}
\caption{
Structure functions of the {\it Swift}/XRT count rate at order $n=2,4,6,8$
   (red, yellow, green and blue curves, respectively)
   for the light-curve interval
   $4 \times 10^5 {\rm s} < t < 4.7 \times 10^6$~s.
The thickness of each curve indicates $1\sigma$ uncertainties.
Top panel: result based on the direct XRT count rate, 
   without correcting for the long-term decline. 
   We interpret the rising trend towards the longest timescales 
   ($\tau \ga 10^6$~s) as an artefact arising from
   the long-term dimming in the X-ray luminosity. 
Middle panel: result when the count rate is de-trended.
   Locally there is an $e$-folding timescale of $2\times10^6$~s. 
Bottom panel: same as the middle panel, but plotted in logarithmic scale, 
   to highlight the essentially featureless ``red noise''
   at variability timescales $\tau \la 10^5$~s. 
   For $\tau \la 10^5$~s, but longer than the orbital period of {\em Swift} 
($\approx 5800$~s), the structure functions show a featureless power-law form
   $S_n\propto\tau^{n/2}$, indicating stochastic red noise.
}
\label{fig.early}
\end{figure}


\begin{figure}
\begin{center}
\begin{tabular}{cc}
\hspace{-0.6cm} \includegraphics[width=9cm]{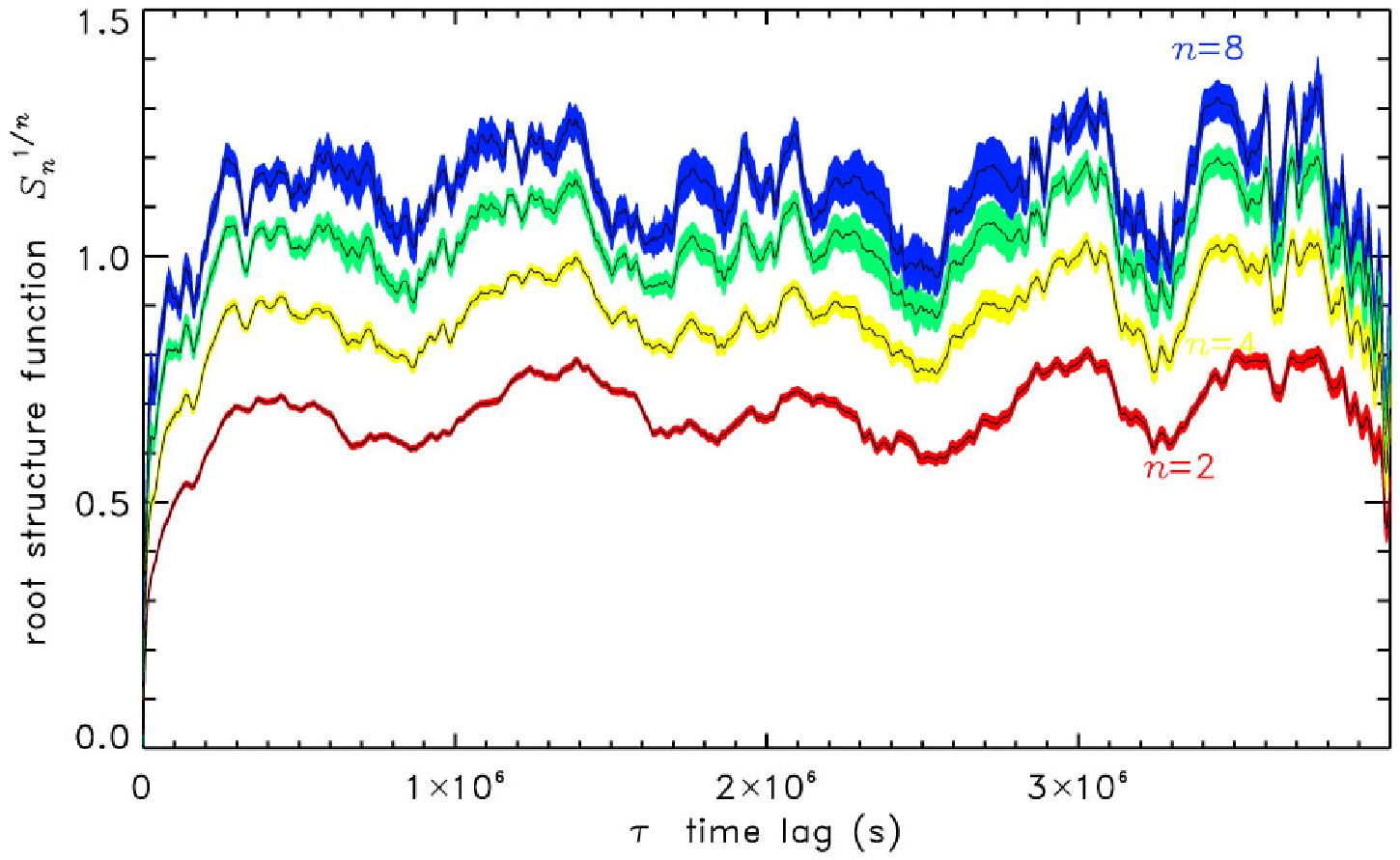}
\\
\hspace{-0.6cm} \includegraphics[width=9cm]{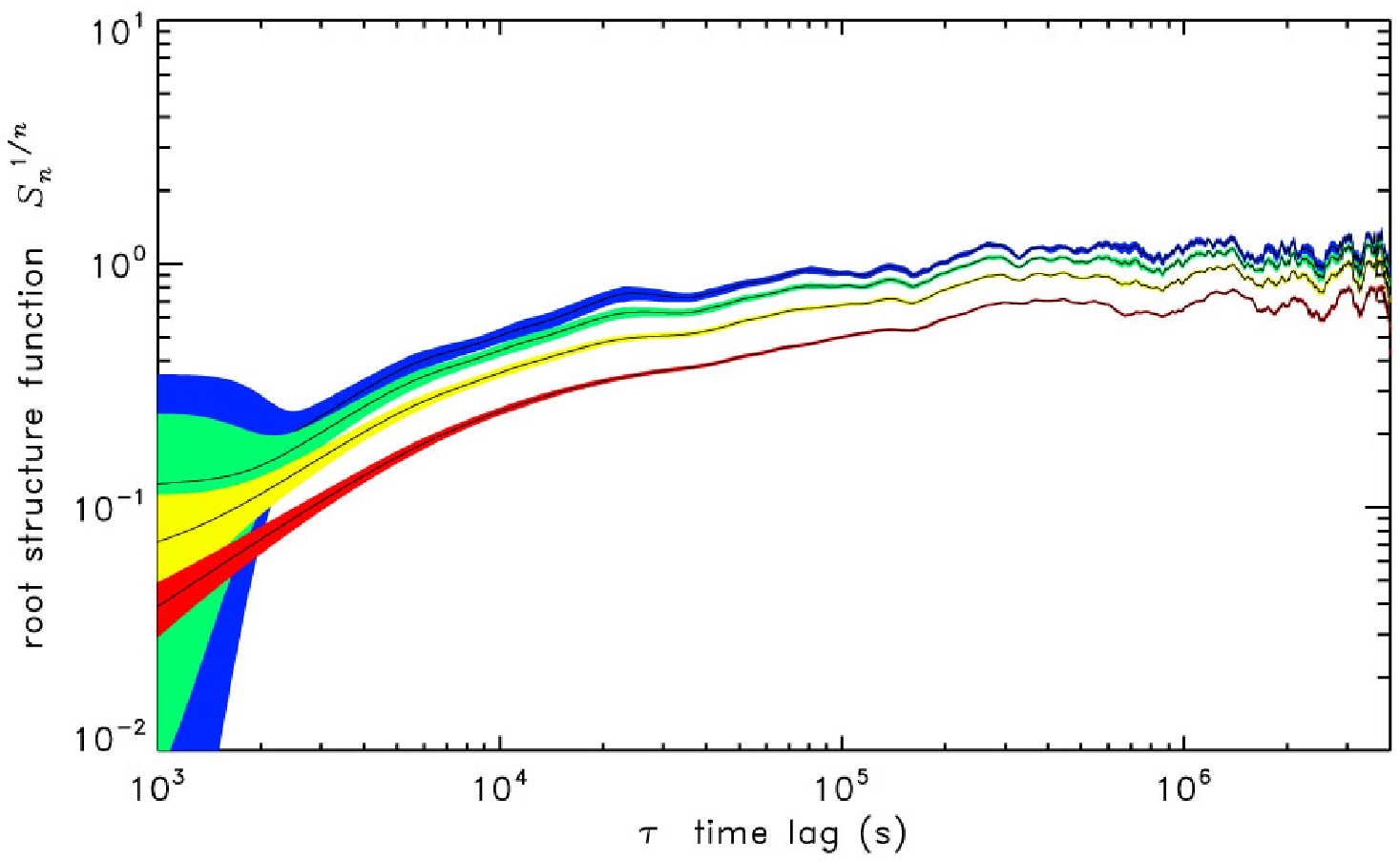}
\end{tabular}
\end{center}
\caption{
Top panel: structure functions of the de-trended X-ray count rate 
at order $n=2,4,6,8$
   (red, yellow, green and blue curves, respectively)
   during the time interval $4.7 \times 10^6 {\rm s} < t < 9.0 \times 10^6$~s.
A period at $\tau\approx 9 \times 10^5$~s (and its multiples)
   is the dominant feature.
The shorter periods that dominated earlier stages of the light-curve
(Figure~\ref{fig.early}) are only slightly visible now,
   and only in higher-order structure functions.
Bottom panel: structure functions in logarithmic scale, showing the white 
and red noise at small values of $\tau$.
}
\label{fig.late}
\end{figure}


\begin{figure}
\begin{center}
\begin{tabular}{cc}
\hspace{-0.6cm} \includegraphics[width=9cm]{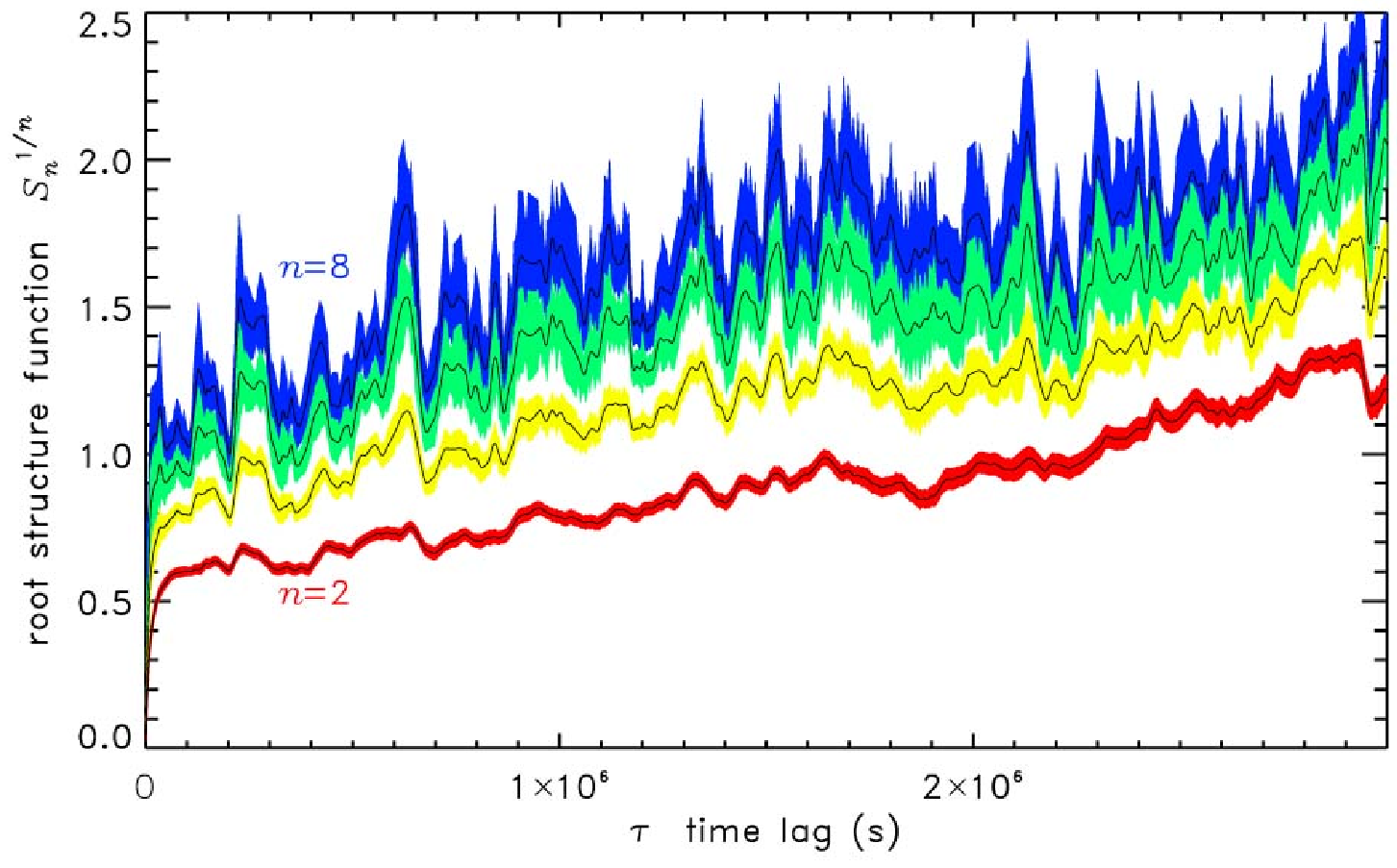}
\\
\hspace{-0.6cm} \includegraphics[width=9cm]{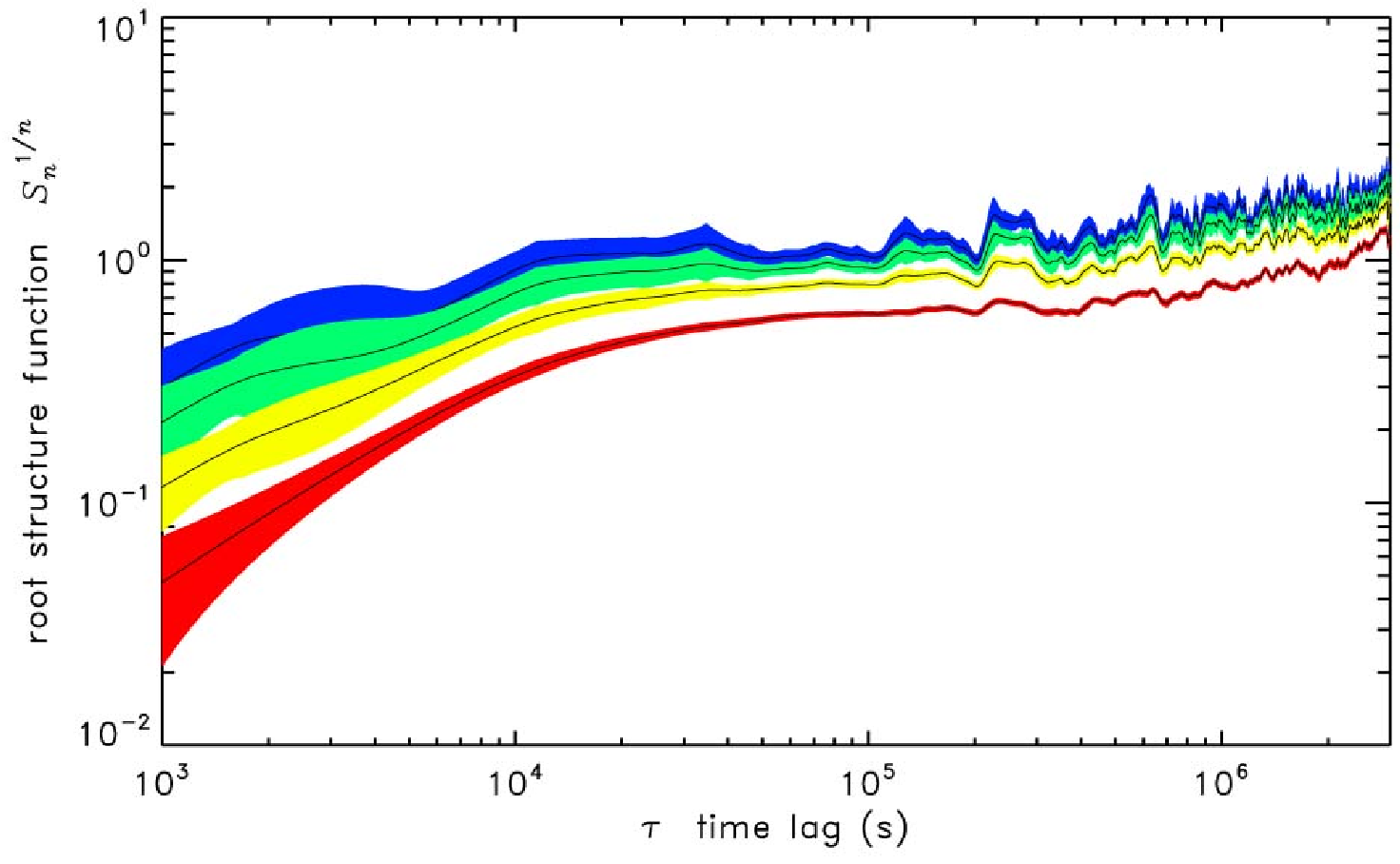} 
\end{tabular}
\end{center}
\caption{
As in Figure~\ref{fig.late}, for 
$9.0 \times 10^6 {\rm s} < t < 12.5 \times 10^6$~s.
}
\label{fig.plateau}
\end{figure}

\begin{figure}
\begin{center}
\begin{tabular}{cc}
\hspace{-0.6cm} \includegraphics[width=9cm]{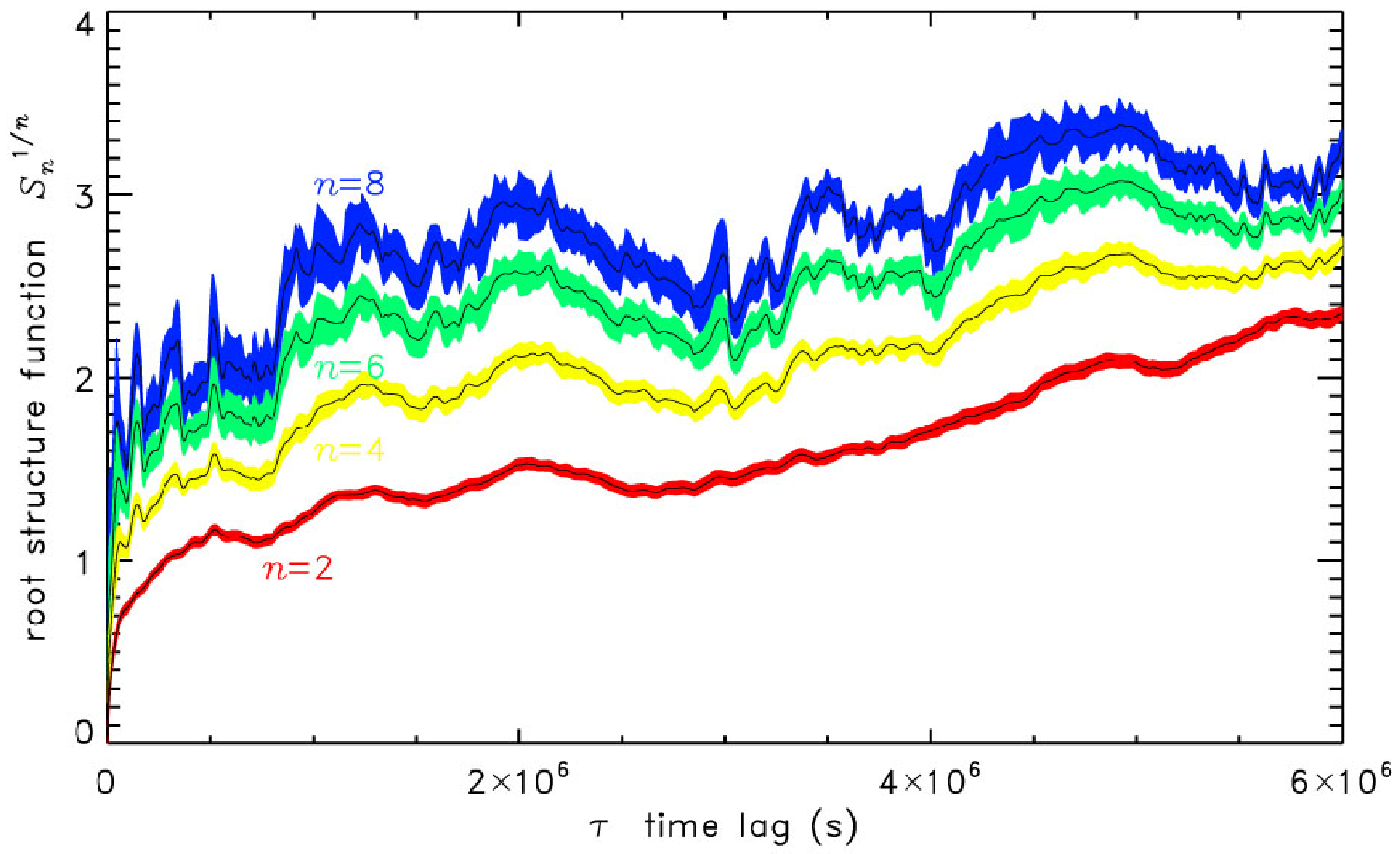}
\\
\hspace{-0.6cm} \includegraphics[width=9cm]{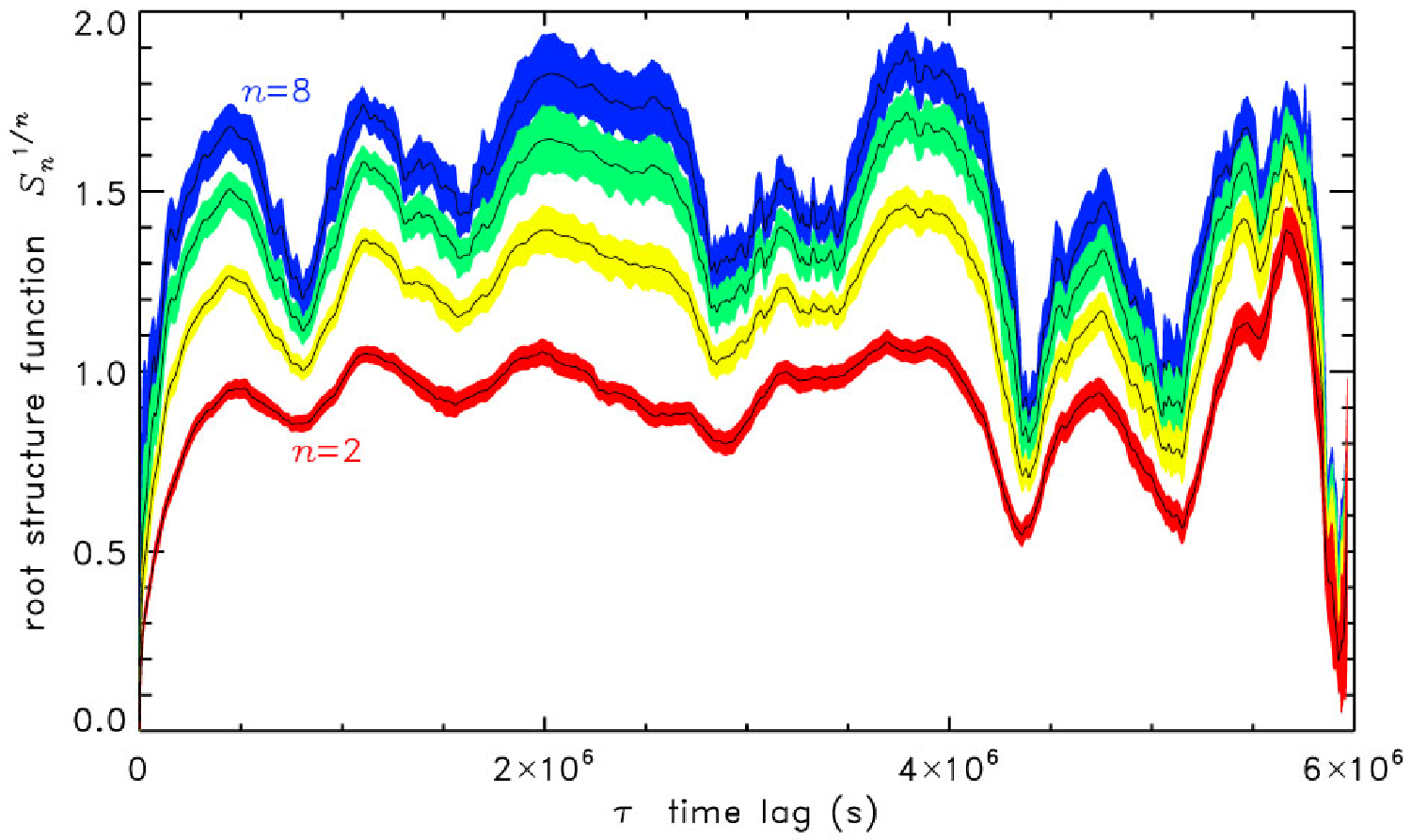}
\end{tabular}
\end{center}
\caption{
Structure functions of the detrended X-ray light-curve,
   for the recent epoch following the end of the plateau stage.
The top panel is integrated over the whole available sequence,
   ($t > 1.25\times10^7$~s).
The bottom panel uses only a shorter interval
   ($1.25\times10^7<t<18.5\times10^7$~s)
   before the observations became too sparse.
}
\label{fig.recent}
\end{figure}

As the X-ray light-curve is not a simple exponential or power-law decline, 
   but seems to go through phases of different slope, 
   we calculated separate structure functions during different time intervals, 
   to determine whether the charactieristic variability timescale
   changed in different epochs.
We calculated two sets of functions:
   one with the count rates directly detected by the XRT;
   the other with a de-trended count rate,  
   to remove the long-term evolution (timescale of weeks/months) 
   and highlight the short-term variability (timescale of a few days).
At early epochs (in particular, at times 
   $2.0\times10^5 {\rm s} \la t \la 9.0\times10^6$~s; Figure 1),
   multiplying the count rate by an exponential function $\exp(t/t_0)$,  
   with $t_0 \approx 2.0 \times 10^6$~s,
   is a simple way to correct for the long-term evolution.
For the whole light-curve 
   (including the later plateau stage), a simple exponential 
   or power-law function is not applicable;
   we used empirically fitted rational functions.
Note that the results that we obtained are independent of
   the precise choice of normalising function for the count rate, 
   and including or omitting data from the first few days
   also does not significantly alter the outcome. 
Our choice of binning (for example, 
   using the same binning of the snapshot observations,
   or a fixed time binning of 100 or 200 s)
   does not have significant effects, either.


What we are looking for in a structure function $S_n(\tau)$
   are local depressions around specific $\tau$ values: 
   they indicate that the light-curve is repetitive on that period.
We obtain best estimates of the periods
   by numerically locating the minima in $S_2$
   (the smoothest structure function)
   and then discarding all those that are shallower than
   the local uncertainty $1\Delta S_2$ ($1\sigma$ result)
   or $3\Delta S_2$ ($3\sigma$ result).
We discard the lesser minima that are within 
   the $1\sigma$ ($3\sigma$)
   catchments of wider and deeper minima.
Table~\ref{table.dips1}
   shows results obtained from data spanning the decaying phase
   of the light-curve.

\subsubsection{Variability during the decay phase}

The most significant result of our analysis is that 
   for $t \la 9 \times 10^6$~s, all structure functions 
   exhibit strong minima at integer and half-integer multiples 
   of a basic period $\tau \approx 4.5 \times 10^5$~s
   (see Figure~\ref{fig.early} 
   for the very early epoch,
   and Figure~\ref{fig.late} for a later epoch).
Most of these features are many times deeper
   than the sizes of the statistical uncertainties.
Corresponding features appear in structure functions of different order.
There is also a significant difference between structure functions calculated 
   for early epochs ($t \la 4.7\times 10^6$s)
   and later in the decline phase 
   ($4.7\times 10^6$~s $\la t \la 9.0\times 10^6$~s).
At early epochs, 
   the most prominent feature is in fact at
   $\tau \approx 2.3 \times 10^5$~s, 
   that is half of what we identified
   as some kind of fundamental timescale.
Figure~\ref{fig.early}
   shows 17 consecutive features in the structure functions,
   at integer multiples of $2.3 \times 10^5$~s.
At later times, the shorter periods ($\tau \approx 2.3 \times 10^5$~s 
   and $\tau \approx 4.5 \times 10^5$~s)
   fade and remain only marginally 
   detectable in the higher order structure functions.
Instead, 
   the deepest features in the structure functions are at
   $\tau \approx (0.9, 1.8, 2.7, 3.6) \times 10^6$~s,
   higher multiples of the fundamental timescale (Figure~\ref{fig.late}).

The characteristic timescales found from structure function
analysis (in particular, the features at $2.3 \times 10^5$ s, 
$4.5 \times 10^5$~s and $9 \times 10^5$~s) confirm and strengthen the results 
obtained from the Lomb-Scargle periodogram (\S~\ref{s.lomb})
and are consistent with a visual inspection of the lightcurve 
(Figure~\ref{fig.first_dips}).
The two shortest frequencies correspond also 
to the two features we identified as marginally significant 
in the power density spectrum (\S~\ref{s.fourier}).
None of these timescales appear to be exact periodicities:  
   sometimes a major dip may commence slightly earlier or later
   than this time interval,
   or may not appear at all,
   but subsequent dips tend to return to the original phase
   (see intervals marked in the top panel of Figure~\ref{fig.first_dips}).

\subsubsection{Variability during the plateau phase}

After $t \approx 9 \times 10^6$~s, the slope of the long-term trend 
   flattened and the light-curve settled on a plateau (Figure~\ref{fig.lightcurve}),
   with an X-ray count rate $\approx 0.1$ ct s$^{-1}$.
At the same time, major dips,
   which had punctuated the light-curve until then,
   appeared to vanish.
Structure functions calculated for the plateau stage
   do not show any significant periodicity
   (Figure~\ref{fig.plateau}).
Depending on the specific choice of detrending function,
   the fluctuations are ``white noise'' (flat shape)
   or flattish ``red noise'' (rising at long $\tau$)
   for timescales $\tau\ga 5 \times 10^4$~s.
As always, at shorter timescales
   the slopes of the structure functions indicates red noise
   down to the orbital period of {\em Swift}
   and the typical gap interval between observations.

\subsubsection{Variability during later epochs}

A dipping behaviour began again at $t\approx1.25\times 10^7$~s.
The range of count-rate variability is about the same 
as in the early phases, that is a factor $\approx 10$--$20$ 
between peaks and troughs, but the dips are broader. 
The dip duration has become comparable to the inter-dip duration.
In the early epochs, 
the typical duration of a dip is $\sim 1$ day; at later epochs, 
it is $\sim 10$~d (cf.~top and bottom panel of Figure~\ref{fig.first_dips}).

Figure~\ref{fig.recent}
   shows the structure functions of the detrended light-curve
   during the most recent epoch.
There are minor wiggles at 0.1Ms timescales.
These are probably spurious,
   as they are comparable to the gaps between observations.
When the whole time interval $t\ga1.25\times 10^7$~s 
is included in the structure function calculation
   (top panel of Figure~\ref{fig.recent}), 
   it is not immediately obvious whether 
there are dominant timescales. 
This is partly because the {\it Swift}
   monitoring observations have become less frequent,
   and partly because the source has not varied much after 
   $t\approx1.85\times 10^7$~s.
Taking only the sub-interval $1.25\times10^7 < t < 1.85\times10^7$~s, 
when the source is more variable, reveals clearer features at
   $\tau\approx0.7, 1.4, 2.9, 4.3, 5.0$ and $5.7$~Ms
(bottom panel of Figure~\ref{fig.recent})
   --- all simple multiples of the same basic frequency,
   although it is a different base frequency than the one 
   associated with the narrow dips in the initial rapidly declining stages.
The variability timescale $\tau\approx1.4\times10^6$~s
   was also identified at high significance in the Lomb-Scargle periodograms 
   (fourth panel of Figure~\ref{fig.lomb}).
In fact, it appears more significant in the Lomb-Scargle periodograms
   than in the structure functions.
This is probably because of the nearly 
   sinusoidal waveform of the recent oscillations.

\subsection{Time-dependent spectral properties} 

\subsubsection{Hardness ratio evolution}
\label{s.hardness.evolution}

\begin{figure}
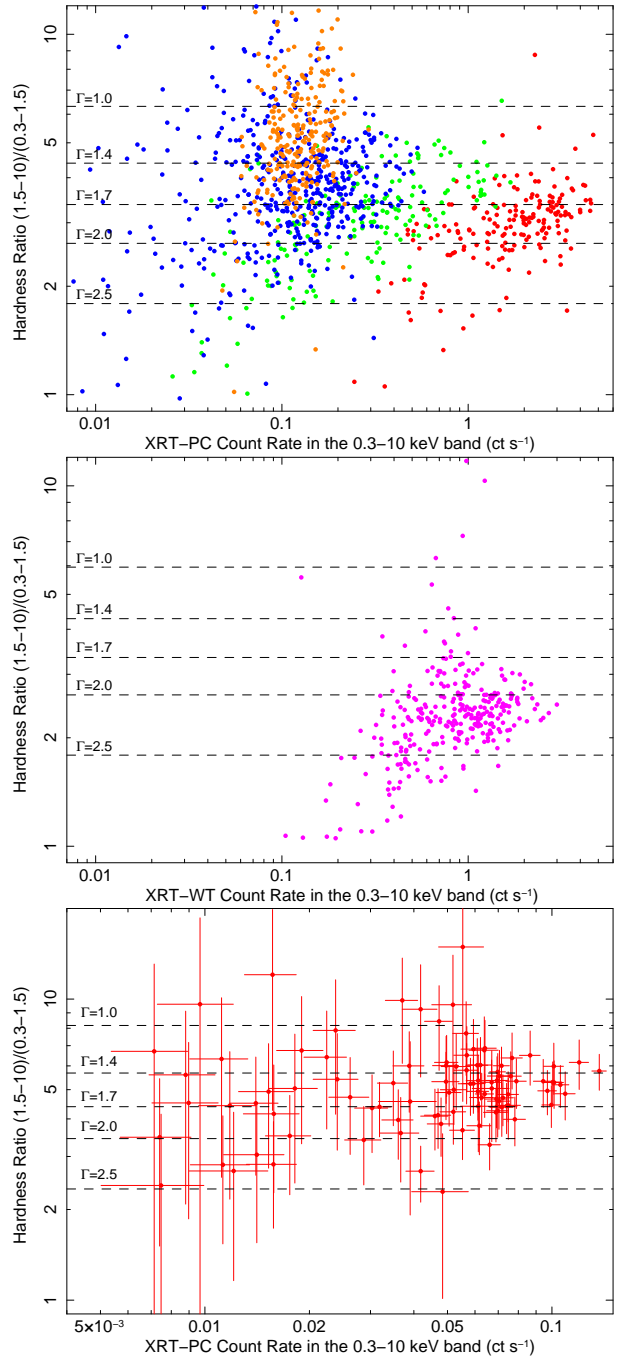

\begin{center}
\includegraphics[height=80mm,angle=270]{hardness_snap_pc.ps}\\
\includegraphics[height=80mm,angle=270]{hardness_snap_wt.ps}\\
\includegraphics[height=80mm,angle=270]{hardness_obs_pc.ps}
\end{center}
\caption{Top panel: hardness-intensity diagram for the snapshot datapoints 
in PC mode. Red datapoints are those for $t < 1.25 \times 10^6$~s; 
green datapoints for $3.41 \times 10^6 < t < 4.7 \times 10^6$~s; 
blue datapoints for $4.7 \times 10^6 < t < 9 \times 10^6$~s;
orange datapoints for $9 \times 10^6 < t < 1.25 \times 10^7$~s. 
Error bars have been omitted for clarity.
The horizontal dashed lines correspond to the hardness ratio 
expected in PC mode for a power-law spectrum with intrinsic column density 
$N_{\rm H} = 1.5 \times 10^{22}$ cm$^{-2}$ (at redshift 0.354), a Galactic 
absorption $N_{\rm H} = 2 \times 10^{20}$ cm$^{-2}$, and several different 
values of the photon index $\Gamma$.
Middle panel: hardness-intensity diagram for the datapoints in WT model, 
at time $1.25 \times 10^6 < t < 3.41 \times 10^6$~s, with intrinsic 
$N_{\rm H} = 1.5 \times 10^{22}$ cm$^{-2}$.
Bottom panel: hardness-intensity diagram in PC mode for the last part 
of the light-curve ($1.25 \times 10^7 < t < 2.1 \times 10^7$~s), 
binned to observation datapoints; an intrinsic 
$N_{\rm H} = 1.9 \times 10^{22}$ cm$^{-2}$ was used for this late epoch, 
as suggested by our spectral fitting (see \S~\ref{s.spectral}).
}
\label{f9}
\label{fig.hard1}
\end{figure}

It was noted in previous work \citep{bur11,lev11b}
   that the X-ray emission of Sw~J1644$+$57
   is harder (flatter spectrum) 
   when the source is brighter.
However, this one-dimensional correlation cannot fully 
characterise the spectral behaviour.
The count rate 
   drops during the short-term dips,
   but also as a result of the long-term evolution:
   we need to treat the two effects separately, 
   as they probably have different physical origins.
To do so, we divided the light-curve into several time intervals, 
   similarly to our approach with the Lomb-Scargle periodograms 
and structure function analysis; 
the intervals are also defined in approximately the same way for convenience, 
although the precise start and end of each phase is somewhat arbitrary. 
We plotted the hardness ratio (defined as the observed count rate 
   in the $1.5$--$10$~keV band over that in the $0.3$--$1.5$~keV band) 
   as a function of total count rate for each segment, in different colours,
   in Figure~\ref{fig.hard1}. 

At early times, the general trend is a moderate hardening 
as the baseline flux declines, alternating with significant softening 
during each dipping episode. This is particularly evident in the 
distribution of datapoints coded with red, green and magenta colours 
in the top two panels of Figure~\ref{fig.hard1}, which 
cover the time interval $2 \times 10^5 < t < 4.7 \times 10^6$~s.
Further independent confirmation of the softening behaviour 
   during the dips comes from {\it XMM-Newton} observations 
   taken on 2011 April 16, April 30, May 16 and May 30. 
The power-law photon indices measured during the first, 
   second and fourth observation were
   $\Gamma \approx 1.88, 1.66, 1.61$ 
   respectively \citep{mil-stro2011},
   consistent with the moderate 
   long-term hardening as the baseline flux declined; 
but the third observation occurred during a dip, and the photon index 
was $\Gamma \approx 1.97$.

At later times,
   as the light-curve reaches a plateau and dips become less frequent,
   the hardness ratio also seems to settle 
   around a constant value corresponding to $\Gamma \approx 1.4$, 
   with some scatter mostly due to the small 
   number of counts in each snapshot observation. 
When the dipping behaviour resumes, at time $t \approx 1.25 \times 10^7$~s,
   the hardness ratio no longer changes during the dips 
(bottom panel of Figure~\ref{fig.hard1}), 
   in contrast with the behaviour at early times.

\begin{figure}
\begin{center}
\includegraphics[width=84mm]{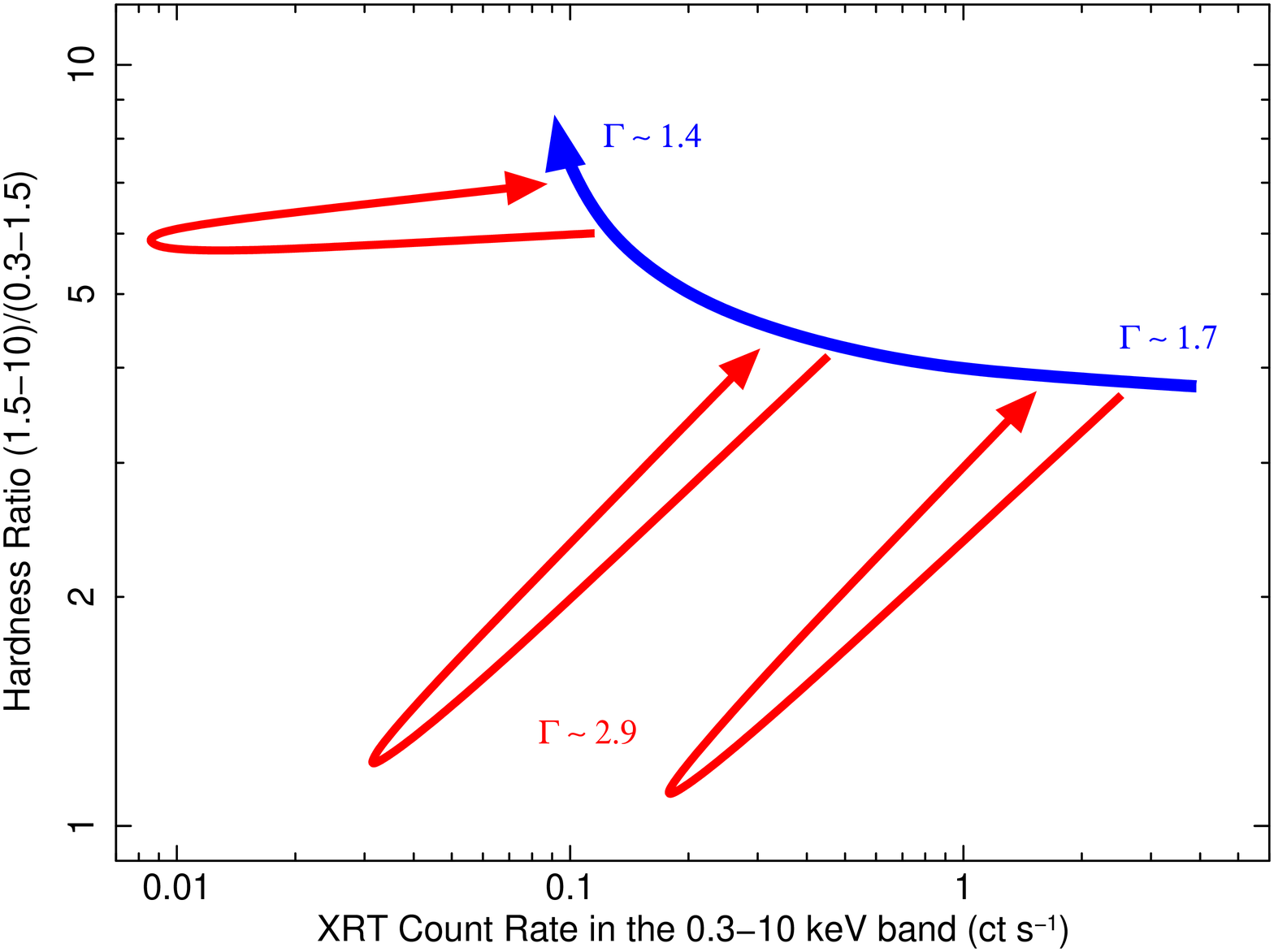}
\end{center}
\caption{Schematic interpretation of the hardness-intensity plots.
The fast decline during the first 3 months is associated to a moderate 
hardening (from $\Gamma \approx 1.7$--$1.8$ to $\Gamma \approx 1.4$) 
followed by a plateau. At early stages, the spectrum becomes much softer 
during each dip; at later times, the hardness ratio does not change 
in the dips.}
\label{f4}
\label{fig.sketch}
\end{figure}

In summary, we found that at early epochs, the dips are shorter, sharper 
and significantly softer than the baseline emission at the same epoch. 
At late times, instead, the dips become broader and longer, and 
the hardness ratio is independent of flux.
We sketch our interpretation of the flux-hardness evolution 
in Figure~\ref{fig.sketch}.


\begin{table*}
\begin{center}
\begin{tabular}{lrrrrr}
\hline
Parameter & {Band 1 Value} & {Band 2 Value} & {Band 3 Value} 
& {Band 4 Value} & {Band 5 Value}\\
\hline
\multicolumn{6}{c}{Model: {\tt tbabs*ztbabs*power-law}} \\
\hline\\[-5pt]
$N_{\rm H,Gal}$ & $[0.02]$ & $[0.02]$ & $[0.02]$ & $[0.02]$ & $[0.02]$\\[5pt]
$N_{{\rm H},z=0.354}$ & $1.57^{+0.10}_{-0.10}$ 
         & $1.41^{+0.09}_{-0.09}$ 
         & $1.35^{+0.06}_{-0.06}$
         & $1.32^{+0.14}_{-0.12}$
         & $1.46^{+0.44}_{-0.35}$ \\[5pt]
$\Gamma$ & $1.67^{+0.06}_{-0.06}$
         & $1.72^{+0.06}_{-0.06}$ 
         & $1.69^{+0.04}_{-0.04}$ 
         & $1.82^{+0.09}_{-0.09}$
         & $2.87^{+0.47}_{-0.40}$ \\[5pt]
$N_{\rm po}$ & $16.6^{+1.4}_{-1.2}$
         &  $12.0^{+1.0}_{-0.9}$
         & $8.2^{+0.4}_{-0.4}$ 
         & $3.5^{+0.4}_{-0.4}$
         & $1.8^{+1.1}_{-0.6}$ \\[5pt]
\hline\\[-5pt]
$f_{0.3-10}$ & $7.7^{+0.1}_{-0.2}$ 
        & $5.3^{+0.1}_{-0.1}$ 
        & $3.8^{+0.1}_{-0.1}$ 
        & $1.3^{+0.1}_{-0.1}$
        & $0.20^{+0.02}_{-0.04}$\\[5pt]
$L_{0.3-10}$ & $46.8^{+1.3}_{-2.4}$ 
        & $32.4^{+1.4}_{-1.3}$
        & $22.5^{+0.6}_{-0.4}$
        & $8.9^{+0.6}_{-0.5}$
        & $5.8^{+4.0}_{-2.8}$\\[5pt]
\hline\\[-5pt]
$\chi^2_{\nu}$ &  $1.06$ & $0.96$ & $0.96$ & $0.98$ & $1.03$\\
               & $(328.3/311)$ & $(271.1/282)$ & $(461.1/480)$ 
        & $(155.2/158)$ & $(26.8/26)$\\[5pt]
\hline 
\end{tabular} 
\end{center}
\caption{Best-fitting spectral parameters for de-trended 
intensity-selected {\it Swift}/XRT-PC spectra.
Band 1 is for time intervals when the count rate was $> 9/10$ 
of the spline-fitted baseline rate (Figure~\ref{fig.lightcurve});
band 2 for count rates between $2/3$ and $9/10$; 
band 3 for count rates between $1/3$ and $2/3$; 
band 4 for count rates between $1/10$ and $1/3$;
band 5 for count rates $< 1/10$ of the baseline.
All bands include only intervals 
at $2 \times 10^5~{\rm s} < t < 9\times 10^6$~s.
Units: $N_{\rm H,Gal}$
and $N_{{\rm H},z=0.354}$ are in units of $10^{22}$~cm$^{-2}$; 
$N_{\rm po}$ in $10^{-3}$~photons~keV$^{-1}$ cm$^{-2}$~s$^{-1}$ at 1~keV; 
$f_{0.3-10}$ in $10^{-11}$~erg~cm$^{-2}$~s$^{-1}$; 
$L_{0.3-10}$ in $10^{45}$~erg~s$^{-1}$.
Errors indicate the 90\% confidence interval for each parameter of interest.}
\label{tab2}
\end{table*}



\begin{table*}
\begin{center}
\begin{tabular}{lrrrr}
\hline
Parameter & {Band A Value} & {Band B Value} & {Band C Value} & {Band D Value}\\
\hline
\multicolumn{5}{c}{Model: {\tt tbabs*ztbabs*power-law}} \\
\hline\\[-5pt]
$N_{\rm H,Gal}$ & $[0.02]$ & $[0.02]$ & $[0.02]$ & $[0.02]$\\[5pt]
$N_{{\rm H,z}=0.354}$ & $1.50^{+0.07}_{-0.09}$ 
         & $1.54^{+0.19}_{-0.20}$ 
         & $1.79^{+0.06}_{-0.09}$
         & $1.92^{+0.39}_{-0.34}$\\[5pt]
$\Gamma$ & $1.77^{+0.05}_{-0.05}$
         & $1.41^{+0.11}_{-0.11}$ 
         & $1.40^{+0.03}_{-0.04}$ 
         & $1.50^{+0.18}_{-0.17}$\\[5pt]
$N_{\rm po}$ & $30.2^{+1.8}_{-1.6}$
         &  $1.6^{+0.3}_{-0.2}$
         & $1.3^{+0.1}_{-0.1}$ 
         & $1.0^{+0.3}_{-0.2}$\\[5pt]
\hline\\[-5pt]
$f_{0.3-10}$ & $12.2^{+0.1}_{-0.2}$ 
        & $1.1^{+0.1}_{-0.1}$ 
        & $0.91^{+0.05}_{-0.05}$ 
        & $0.54^{+0.04}_{-0.07}$\\[5pt]
$L_{0.3-10}$ & $80.8^{+2.7}_{-2.9}$ 
        & $5.7^{+0.3}_{-0.3}$
        & $4.5^{+0.1}_{-0.1}$
        & $3.2^{+0.2}_{-0.2}$\\[5pt]
\hline\\[-5pt]
$\chi^2_{\nu}$ &  $1.06$ & $0.90$ & $0.92$ & $1.06$\\
               & $(432.3/408)$ & $(95.6/106)$ & $(487.5/528)$ 
                & $(78.4/74)$\\[5pt]
\hline 
\end{tabular} 
\end{center}
\caption{Best-fitting spectral parameters for time-selected 
{\it Swift}/XRT-PC spectra. All spectra include only intervals 
in which the count rate was $> 2/3$ of the normalized continuum rate.
Band A is for time intervals $2 \times 10^5~{\rm s} < t < 4.7\times 10^6$~s; 
band B for $4.7 \times 10^6~{\rm s} < t < 9\times 10^6$~s; 
band C for $9 \times 10^6~{\rm s} < t < 1.25\times 10^7$~s; 
band D for $1.25 \times 10^7~{\rm s} < t < 2.1\times 10^7$~s;
Units are as in Table~\ref{tab1}.
Errors indicate the 90\% confidence interval for each parameter of interest.}
\label{tab3}
\end{table*}

\subsubsection{Spectral evolution} 
\label{s.spectral}

Our investigation of hardness evolution makes it clear that 
   we need to distinguish between the short-term and long-term 
   spectral evolution.
To do so, we use the same method applied 
   to our structure function analysis:
   we normalised the snapshot light-curve
   to a smooth empirical function that fits 
   the long-term evolution.
We then extracted intensity-selected spectra based on the normalised 
   count rates rather than the observed ones; for example,
   integrated spectra from all time intervals
   in which the normalised flux was $> 0.9$, or 
   between $2/3$ and $0.9$, etc.
We also fixed the flux level 
   and extracted time-selected spectra,
   that is with the same normalised 
   flux band but at different epochs.

\begin{figure}
\begin{center}
\includegraphics[height=84mm,angle=270]{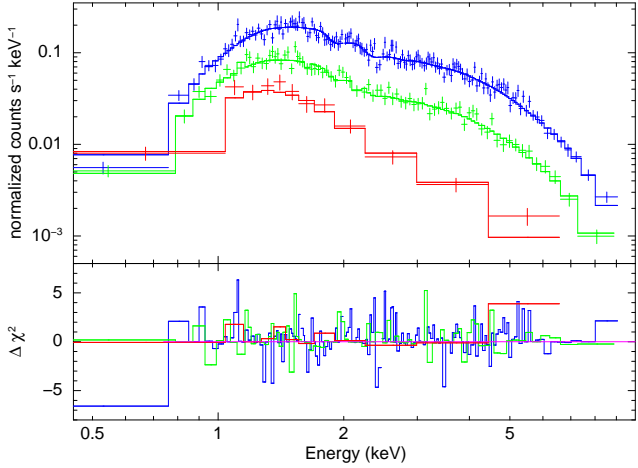}
\end{center}
\caption{Best-fitting power-law models and $\chi^2$ residuals 
for three de-trended intensity-selected spectra, showing 
spectral steepening in the dips. 
Blue datapoints and residuals: background-subtracted 
spectrum extracted from all the time intervals when the count rate 
was $> 9/10$ times the de-trended baseline (Band 1~in Table~\ref{tab2}).
Green datapoints and residuals: spectrum from the time 
intervals when the count rate was between $1/10$ and $3/10$ of 
the de-trended baseline (Band~4 in Table~\ref{tab2}). 
Red datapoints and residuals: spectrum from the time 
intervals when the count rate was $\le 0.1$ times the de-trended baseline 
(Band~5 in Table~\ref{tab2}).
For all three spectra, we used only data up to $t=9\times 10^6$~s;
we used only data in PC mode, to reduce background contamination.
See Table~\ref{tab2} for the fit parameters.}
\label{f11}
\end{figure}

In Table~\ref{tab2}, we summarise the result of spectral fitting 
for a sample of intensity-selected spectra from the decline phase 
($t < 9 \times 10^6$ s).
All spectra are well fitted by absorbed power-laws,
   with a column density 
   $\sim 1.5 \times 10^{22}$~cm$^{-2}$
   in the local frame at redshift 0.354.
Small variations in the fitted column density from band to band 
   may not be statistically significant,
   and do not appear to follow any trend.
The high- and medium-intensity bands have a similar photon index $\approx 1.7$.
Low-intensity bands have a steeper slope,
   as already suggested by our hardness ratio analysis:
   as steep as $\Gamma \approx 2.9$
   for spectra taken across the bottom of the dips (Figure~\ref{f11}).
This result confirms our interpretation of the hardness-intensity plots 
and is consistent with the spectral results of \citet{mil-stro2011}. 
It shows even more clearly that
   the early dips are characterised by a steeper spectral slope,
   not by higher absorption:
   they are not eclipses by orbiting or rotating clouds.

\begin{figure}
\begin{center}
\includegraphics[height=84mm,angle=270]{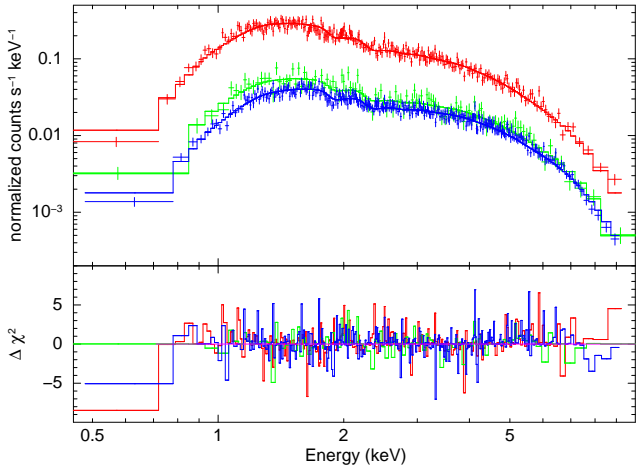}
\end{center}
\caption{Best-fitting power-law models and $\chi^2$ residuals 
for three time-selected spectra, showing moderate long-term hardening. 
Red datapoints and residuals: background-subtracted spectrum extracted 
from the time interval $2 \times 10^5~{\rm s} < t < 4.7\times 10^6$~s 
and only including intervals with a count rate $> 2/3$ times 
the de-trended baseline. 
Green datapoints and residuals: background-subtracted spectrum at 
$4.7 \times 10^6~{\rm s} < t < 9\times 10^6$~s and count rate $> 2/3$.
Blue datapoints and residuals: background-subtracted spectrum at 
$9 \times 10^6~{\rm s} < t < 1.25\times 10^7$~s and count rate $> 2/3$.
All data are in PC mode. See Table~\ref{tab3} for the fit parameters.}
\label{f12}
\end{figure}

We then considered high- and medium-intensity spectra
   (which, as we have seen from Table~\ref{tab2}, 
   have the same average spectral slope),
   taken from all time intervals with normalised count rates $> 2/3$,
   and divided them into several time bands corresponding 
to the intervals used for timing analysis.
We found (Table~\ref{tab3}) that all spectra are well fitted by a simple 
absorbed power-law; adding blackbody or optically thin thermal components 
does not improve the fit, nor is there any evidence of breaks 
in the power-law at high energies.
The slope in the early epoch ($t < 4.7 \times 10^6$ s) 
was $\Gamma \approx 1.8$, flattening to $\Gamma \approx 1.4$--$1.5$ 
at later epochs (Figure~\ref{f12}). This long-term spectral trend is in perfect 
   agreement with what was inferred from the hardness-intensity study 
   (\S~\ref{s.hardness.evolution} and Figure~\ref{fig.sketch}).
The intrinsic column density appears to increase at later epochs 
(Table~\ref{tab3}), although this result is only marginally significant.

At late epochs, the unabsorbed isotropic $0.3$--$10$ keV luminosity 
becomes as low as $\approx 4 \times 10^{44}$ erg s$^{-1}$ 
during the dips (Table~\ref{tab4}). This is now approaching 
the pre-burst upper limit to the X-ray luminosity, 
$L_{\rm X} \la 1.7 \times 10^{44}$ erg s$^{-1}$ \citep{blo11b}.
If Sw~J1644$+$57 continues its decline,
   it is plausible that 
   in a few months' time we will start seeing the unbeamed component
   of the emission
   (e.g., from an accretion disk or its corona), 
   or we will infer stronger limits on the unbeamed flux.
We tried fitting the late-epoch dip spectrum with a thermal-plasma 
model and found that it is at least as good as a power-law model 
   (Table~\ref{tab4});
   there are not enough counts to discriminate between them.


\begin{table}
\begin{center}
\begin{tabular}{lrr}
\hline
Parameter & {Value 1} & {Value 2} \\
\hline
\multicolumn{3}{c}{Model 1: {\tt tbabs*ztbabs*power-law}} \\
\multicolumn{3}{c}{Model 2: {\tt tbabs*ztbabs*raymond-smith}} \\
\hline\\[-5pt]
$N_{\rm H,Gal}$ & $[0.02]$ & $[0.02]$ \\[5pt]
$N_{{\rm H,z}=0.354}$ & $1.8^{+1.6}_{-1.3}$ & $1.7^{+1.2}_{-0.9}$\\[5pt]
$\Gamma$ & $1.58^{+0.73}_{-0.70}$ & -- \\[5pt]
$N_{\rm po}$ & $0.14^{+0.07}_{-0.07}$ & -- \\[5pt]
$kT_{\rm e}$ & -- & $> 3.9$\\[5pt]
$N_{\rm rs}$ & -- & $7.8^{+5.0}_{-1.9}$\\[5pt]
\hline\\[-5pt]
$f_{0.3-10}$ & $0.073^{+0.013}_{-0.012}$ & $0.064^{+0.011}_{-0.032}$\\[5pt]
$L_{0.3-10}$ & $0.45^{+0.30}_{-0.10}$ & $0.39^{+0.08}_{-0.04}$\\[5pt]
\hline\\[-5pt]
Cstat &  $7.35/12$ & $6.16/12$ \\[5pt]
\hline 
\end{tabular} 
\end{center}
\caption{Best-fitting spectral parameters for the {\it Swift}/XRT-PC spectrum 
during the late dips ($1.25 \times 10^7~{\rm s} < t < 2.1\times 10^7$~s), 
fitted with the standard power-law model and with an optically-thin thermal 
plasma model. 
Units of $N_{\rm H}$, flux and luminosity are as in Table~\ref{tab2}.
The normalisation 
$N_{\rm rs} = 10^{-11}(1+z)^2/(4\pi\,D^2)\int(n_{\rm e}n_{\rm I}dV)$ 
where $D$ is the luminosity distance.
Errors indicate the 90\% confidence interval for each parameter of interest.}
\label{tab4}
\end{table}

\section{Discussion} 
  
The general consensus is that Sw~J1644$+$57 is a previously quiescent 
  nuclear BH that underwent an outburst 
  caused by accretion of a tidally disrupted star 
  \citep{blo11a,blo11b,bur11,can11,soc11}.
Its high apparent luminosity requires strong beaming.
The emission in the keV band can be explained 
  as synchrotron and/or inverse Compton radiation
  from energetic electrons streaming along
  the relativistic jet pointing towards us.
The relative contribution of the two processes is still debated.
It is also still unclear how the jets are launched 
  at the onset of the outburst, when a fully formed accretion disc 
  cannot have had the time to develop, after the disruption event 
\citep[cf. the violent mass transfer
	and jet formation in the X-ray binary Cir~X-1;][]{joh99}.
 
From the {\it Swift}/XRT data we have extracted further information 
   on the time-dependent behaviour and spectral evolution of the source. 
We have decomposed the amplitude variations into a long-term evolution 
   (an initial decline followed by a plateau) and recurrent dips.
If these dips had been wholly stochastic processes,
   we would expect that fluctuations should be
   of the same order of magnitude as the depths of the dips.
Otherwise it would require many miniature flares
   (a day's worth of them)
   to coherently conspire to become small simultaneously,
   or to abstain simultaneously.

However we have shown that the dips are not merely random fluctuations 
   (as is often the case in the X-ray light-curve of accreting sources),
   but have a certain regularity,
   apparently governed by an underlying periodic driver.
We have used Lomb-Scargle periodograms and structure function analysis 
to search for the main characteristic timescales 
in the frequency range $\sim 10^{-7}$--$10^{-4}$ Hz.
We expect variability on those timescales based on physical 
arguments. For example, there may be a residual inhomogeneous 
distribution of debris at or near the tidal disruption radius 
$R_{\rm TD} \approx 10GM_\bullet/c^2 \times [M_\bullet/(10^7 M_{\odot})]^{-2/3}$ 
(for a solar-type star). 
The Keplerian timescale $t_{\rm K} = 2\pi/\Omega_{\rm K}$ 
at the tidal disruption radius is $\approx 10^4$ s \citep{ree88}, 
independent of BH mass 
in the Newtonian approximation. The ring of debris 
is likely to be still oriented in the orbital plane of the disrupted 
star: this may generate a Lense-Thirring precession with a timescale
$t_{\rm LT} = [t_{\rm K}/(2a)]\times[(R_{\rm TD}c^2)/(GM_\bullet)]^{3/2} 
\sim 30$--$300 t_{\rm K}/(2a) \sim 10^5$--$10^6$ s 
\citep{armitage99,mer10,sto11}, where $a$ is the BH spin parameter.
We found significant features in the Lomb-Scargle periodograms 
   and structure functions, especially at 
   $\tau \approx 2.3$~Ms, 4.5~Ms, 9~Ms during the early epochs,
   and $\tau \approx 0.7$~Ms and 1.4~Ms 
during the late epochs. It appears that the characteristic dipping  
timescale (and its associated harmonics) has shifted between the early 
and late section of the lightcurve.

Whatever the mechanism for the dips, it is something that also changes 
   the spectral properties of the emission,
   as the spectrum is steeper during the early dips
   ($\Gamma \approx 2.8$--$3$)
   than immediately before and after 
   ($\Gamma \approx 1.4$--$1.7$).
At the same time, the long-term trend 
outside the dips shows spectral hardening at the beginning, 
followed by a plateau at $\Gamma \approx 1.4$.


Here, we attempt to make sense of our findings 
   in the framework of a synchrotron-emitting relativistic jet scenario.   
We propose that: 
(i) the jets have precessional and nutational motion; 
(ii) the jets have a collimated core surrounded by an envelope of 
less energetic, less collimated electrons straying out of the core;
(iii) the synchrotron-emitting jets become more compact
as the source declines, after the initial outburst. 

In this scenario, the baseline flux in the XRT band is direct 
   synchrotron emission from a jet pointing toward us. 
In the first few days after the outburst, when the accretion rate is highest,
   the jet expands and propagates forward into the interstellar medium.
The X-ray emission region is most extended at that early epoch, 
   and it is optically thin to synchrotron radiation.     
Energetic electrons are freshly accelerated in the jets, 
   with an energy distribution $N(E) \propto E^{-p}$,  
   where the energy spectral index
   $2.2\la p\la2.5$
   \citep[{{see e.g.,}}][]{bal92, kir99}.    
The synchrotron emission spectrum from this population of electrons 
   has a specific intensity $I_{\nu} \propto \nu^{-\alpha}$ 
   with $\alpha \approx 0.7$, 
   corresponding to a photon index $\Gamma \approx 1.7$.   
When the most violent flarings subside and accretion becomes more steady,   
  the overall accretion luminosity of the system decreases and the jet power 
  is reduced accordingly.
In this phase, synchrotron emission no longer 
   comes from the expanding jet ejecta, but it is rather confined 
   to the vicinities of shocks at the base of the jet.  
As the X-ray emission region becomes more compact and denser, 
   the synchrotron photons start to suffer self-absorption, 
   leading to the observed spectral flattening:
   approaching a similar situation 
   to the flat-spectrum core emission in radio galaxies with compact jets.

Moreover, the transition from extended to compact jet emission   
   may also provide a simple explanation for the levelling off of the X-ray 
   luminosity at late epochs.   
For an optically thin source, the luminosity depends on the total volume 
   of the synchrotron-emitting region in the jets;
   a constant luminosity would require a high degree of fine-tuning.
By contrast, in an opaque source,  
   the luminosity depends on the effective area
   of the emission region visible to us.     
The growth of the radio emitting region 
   is perhaps explained by a more diffuse outer region,
   such as the larger jet-driven turbulent cocoon
   and the expanding bow shock surrounding the whole system 
   \citep{bow11,zau11,bur11}. 
Modelling the radio flux evolution requires knowledge of the jet parameters,
   which are unfortunately not well constrained
   by the observational data obtained so far.
We therefore leave the jet modelling,
   which is beyond the scope of this paper, to a future study.

Regarding the dips, we showed that they are not due to partial occultation 
   of the jets by some line-of-sight material, 
   as we would expect higher absorption rather than spectral steepening 
   in this situation, contrary to what is observed. 
Instead we propose that dips occur
   when the synchrotron-emitting jet is not perfectly aligned 
   along our line-of-sight, 
   and hence the most highly beamed component of the synchrotron 
   emission does not reach us. 
The observed X-ray flux is then dominated by the lower level of emission 
   from the slower and/or less collimated electrons in
   a sheath or cocoon around the jet. 
This emission could be synchrotron from
   an aged (steeper-spectrum) population of electrons,
   if there are significant magnetic fields in the jet envelope, 
or synchrotron self-Compton.
We must also stress that even at the bottom of the dips, the apparent 
   luminosity is higher or similar (for the late dips) 
   to the Eddington luminosity of a $10^{7} M_{\odot}$ BH. 
This suggests that the emission is moderately beamed even during the dips.

In addition to the expected long-term decline and stochastic flux 
variability, it is intriguing that many of the dips occur on regular patterns,
   likely to be governed by a periodic driver;
   it is even more intriguing that sometimes a dip 
   is skipped but the following one happens again
   approximately at the expected phase
   (Figure~\ref{fig.first_dips}).
Careful inspection of the centroid locations of the dips has revealed 
   that they are not exactly phase-matched but can be slightly advanced 
   or delayed.
To explain these peculiar dip patterns, 
   we propose that the jet undergoes nutation as well as precession. 
The combination of the two effects drives the cone of the jet out of 
   our line of sight, from time to time.  
The presence of nutation is essential, 
   as it naturally explains why the dips occur
   in a regular but not exactly periodic pattern,
   and why there are slight advances or delays of the dip centroids.

We note also that the appearance of the dips changes:
   in the early epoch we see a slow fainting and rapid rise;
   but in later dips the dimming and rising are symmetric
   (Figure~\ref{fig.first_dips}).
The fractional time that the object
   is in a dip or out of a dip also changes:
   brief dips at early times; long dips in late epochs. 
These observations gain significance
   in the jet precession/nutation scenario,
   especially if the emission core became more compact.
When the early dips were sharp and brief,
   the precession subtended a tight angle,
   compared to the jet's beaming angle.
The larger width of the dips at late times,
   and the larger fraction of time spent in a dip,
   probably indicates that 
   the core of the jet is narrower
   (or swinging more widely)
   while the layer of slower electrons around the core grows in size.
The asymmetric dipping at early times may mean that 
   the region around the core is 
   not moving in phase with the jet core;
   lagging while the core moves out of the line of sight,
   while being compacted when the core moves back.

The remaining question is what causes the jet precession and nutation. 
Jet precession is expected as the BH spin axis is unlikely to be perfectly 
   aligned with the initial angular momentum of the disrupted star. 
Thus, the normal direction to the plane of the transient accretion ring/disc 
   formed by the stellar debris is tilted with respect to the BH spin, 
   and this causes the jet to precess.
In addition, jet nutation can be triggered
   if the precessing jet/disc system has been perturbed,  
   either by an internal or external driver.
In the internally driven case, 
    nutation may be induced by an uneven mass distribution in the transient 
    debris ring/disc.
In the externally driven case, 
   the jet and the accretion ring/disc may be perturbed by the orbiting 
   remnant of the partially disrupted star ---
   for example, a white dwarf on an elliptical orbit
   would make several passes inside the tidal 
   radius before being completely disrupted \citep{kro11}.
Alternatively,  
   they may be perturbed by another gravitating object, such as 
   (unlikely but not impossible) a satellite massive black hole
   \citep[as predicted
	in nuclear swarms of compact objects:][]{freitag2006,portegies2006};
   or perhaps a more supermassive companion
   (about which Sw~J1644$+$57 is the satellite).

An important clue to the evolution of the system comes from the change 
   in the characteristic variability timescales at early and late times.
If the nutation is externally driven,
   the change of period might conceivably be explained
   if the torque changed 
   (perhaps during an eccentric perimelasma passage).
Scenarios dominated by disc torques suggest other possibilities.
A possible explanation is that the warped disk has not yet reached 
   a steady state or Bardeen-Petterson regime 
   \citep{bar75,kum85}.
In other words, the warp radius
   (transition between the inner part of the disc 
   aligned with the BH spin and the outer part 
   aligned with the orbital plane of the mass donor)
   may still be propagating outwards,
   and as a result the Lense-Thirring precession timescale 
   at the warp radius is increasing. 
Alternatively, it was suggested \citep{fra07,dex11,sto11} 
   that a geometrically thick disc 
   may never settle in a Bardeen-Petterson regime,
   and may instead precess as a solid body
   (a tilted, rather than warped disc).
In this case, the precession frequency depends on the dimensionless 
   radial surface density profile of the disc, 
   which is also likely to evolve in time.

In fact, a problem of the disc-driven precessing-jet scenario is that 
   we expect the jet axis to have already precessed out of alignment 
   with our line of sight (and not just for short dips) after few weeks, 
   unless the initial alignment of disc and BH spin axis was already 
   very close \citep{sto11}.
This seems an unlikely coincidence 
   if the disc was formed after a tidal disruption event.
Thus, we suggest a note of caution regarding the tidal disruption 
   scenario as the origin of the flare. It is useful to remember 
   that the 2011 March 28 flare marks the moment when the jet turned on,
   not the time of the putative tidal disruption event,
   which may have happened unobserved several weeks earlier,
   depending on the time required to spread 
   the debris onto a disc, build up a magnetic field and launch a jet.
The main argument for a tidal disruption as the origin 
of the whole process is that the host galaxy was not known as an AGN 
before the flare.
However, pre-event luminosity upper limits \citep{blo11b}
   are only implying that the bolometric luminosity
   was below Eddington for a $10^7 M_{\odot}$ BH,
   or below $10$ Eddington for a $10^6 M_{\odot}$ BH.
Thus, we cannot exclude that an accretion disc was already present 
and approximately aligned, and the jet ejection was due to a state transition 
rather than a tidal disruption. State transitions with relativistic blob 
ejections and radio/X-ray flares are common in Galactic BH transients, 
especially when they switch from the hard to the soft (thermal) state 
\citep{fen04,fen-bel04}. Scaled with their respective mass ranges, 
the duration of the flare in Sw~J1644$+$57 so far corresponds to only 
$\sim 10$--$100$ s in a typical Galactic BH. Further monitoring 
of the flux evolution over the next few months will help testing 
between the tidal disruption and state transition scenarios.

The scenario outlined above does not imply that {\it every} dip is due 
to oscillations 
and precession of the jet. Our most general interpretation is that 
dips correspond to phases when we are seeing slower and less collimated 
electrons along our line of sight in the jet. 
Even when the nozzle is steady, jets are subject to internal instabilities
of magneto-hydro-dynamical nature. Internal shocks echo and self-intersect 
up and down the jet, producing standing patterns with a characteristic
spacing that depends on the Mach number of the flow and the width of the jet. 
Jets also suffer an axisymmetric pinch instability, a non-asxisymmetric 
helical instabilities, and other higher-order harmonics of the helical 
instability \citep{har88,nor88,har92}.
Observed changes in the jet collimation and luminosity will depend on 
the interplay of the external driving oscillations ({\it e.g.}, jet 
precession) and the resonant frequencies of the internal instabilities. 
The situation becomes even more complex if --- as seems to be the case 
in Sw~J1644$+$57 \citep{ber11} --- the jet has a broad Lorentz factor 
distribution rather than a single speed.
With more arduously detailed modelling work (in the future), tested 
by long-term X-ray and radio monitoring \citep{ber11},
   one might hope to infer something about
   the internal layering of the jet, and distinguish between 
internal and external driven jet variability.


\section{Conclusion}   
   
We investigated the luminosity and spectral evolution of
   the peculiar X-ray source Sw~J1644$+$57 using {\em Swift}/XRT data. 
Our structure function analysis showed that 
   the large-amplitude variations in the light curve of the source 
   are not simply stochastic fluctuations. 
In particular, the occurrence of at least some of the dips 
   appears to follow some regular patterns 
   characterised by multiples and fractions of a time interval 
   $\tau \approx 4.5 \times 10^{5}$~s;
   visually, several prominent dips 
   are spaced at $\approx 0.5 \tau \approx 2.2 \times 10^5$~s, 
   and $\approx 2 \tau \approx 9 \times 10^5$~s. 
After the plateau the base period switched to
   $\tau\approx1.4\times10^6$~s,
   but the scenario remains essentially the same.
The X-ray spectrum outside the dips
   is always consistent with an absorbed power-law,
   but its photon index evolves on a weekly/monthly timescale, 
   from $\Gamma \approx  1.8$ at the beginning of the decline phase 
   to $\Gamma \approx  1.4$
   at a later stage when the luminosity seems to level off.
The X-ray spectrum is much softer during the dips,
   while there is no significant change in the absorption column density. 

We proposed a scenario in which the synchrotron emitting jets,
   launched from a massive BH accreting a disrupted star, 
   undergo both precession and nutation. 
The baseline flux in the keV band is direct synchrotron emission 
   when the jet cone is in our line of sight;
   dips occur when the jet cone goes briefly and partially 
   out of our line of sight. 
We argued that the synchrotron photons come from the optically thin 
jet ejecta during the initial high-luminosity, exponential decay phase; 
   as the accretion power subsides,
   the dominant emission region becomes
   an opaque feature that is confined to a compact jet core.        
We attributed the fainter, steep-spectrum dip emission to a population 
   of less energetic electrons, perhaps streaming 
   in a cocoon surrounding the collimated jet core. 
Jet nutation and precession provide a natural explanation 
   for the dip patterns in the X-ray light-curve: 
   in particular, their preferred occurrence at some regular intervals, 
   their occasional disappearance and subsequent reappearance, 
   and the phase advance and delay of the dip centroids. 
In addition, internal jet instabilities can produce oscillations 
in the speed and cross section of the jet flow, and therefore 
a recurrent dipping behaviour. Long-term monitoring of the characteristic 
oscillation frequencies will be required to test whether 
they are due to internal or external drivers.


\section*{Acknowledgments}

RS acknowledges support from a Curtin University Senior Research Fellowship, 
and hospitality at the Mullard Space Science Laboratory (UK) and 
at the University of Sydney (Australia) during part of this work. 
This work made use of data supplied by the UK Swift Science Data Centre 
at the University of Leicester. We thank Alister W Graham, Edo Berger 
and the anonymous referee for constructive comments.

\section*{Appendix} 

\subsection*{A.1 Piecewise linear light curves}

Consider a temporally evolving signal $z$
   (e.g. X-ray flux or counts), 
   for which the data occur in bins $i$ spanning temporal intervals
   $[t_i,u_i]$
   where $t_i$ is the starting time
   and $u_i$ is the ending time of bin $i$.
In characterising the temporal variability of the light-curve,
   we may offset a copy of these bins by a lag time $\tau$,
   and compare to the unoffset data.
With patchy, piecewise data, it is necessary to compute the temporal overlap
   between each pair of unoffset and offset bins.
If the time intervals of the offset bins are
   $[t_j^\prime,u_j^\prime]=[t_j+\tau,u_j+\tau]$, 
   then the interval of overlap with an unoffset bin 
   $[t_i,u_i]$
   is
   $[T_{ij},U_{ij}]=[\max(t_i,t_j^\prime),\min(u_i,u_j^\prime)]$.
If $U_{ij}<T_{ij}$, then there is no overlap.
   The weight or duration of overlap between a pair of bins $i,j$ is
\begin{equation}
	w_{ij}	= w_{ij}(\tau)
		= \left\{{
		\begin{array}{ll}
			U_{ij}-T_{ij}	&\mbox{if $U_{ij}\ge T_{ij}$,}\\
			0		&\mbox{otherwise.}
		\end{array}
	}\right.
	\ .
\end{equation}

Suppose that the amplitudes within each bin evolve linearly in time:
   $z=a_i\,t + b_i$
   and
   $z^\prime=a_j\,t + b_j$.
The linear coefficients depend on the (observed)
   amplitudes $z$ and times
   at the ends of each time interval,
   $a_j = ({{\delta z_j})/({u_j-t_j}})$
   and
   $b_j = z_j - a_j t_j$.
%
During the overlap, the difference between the amplitudes
   of offset and unoffset light curves is 
$y \equiv z_j^\prime(t+\tau) - z_i(t) = A_{ij}\lambda + B_{ij}$,
   with $\lambda\equiv(t-T_{ij})/(U_{ij}-T_{ij})$,
   and $0\leq\lambda\leq1$ by construction, 
where we abbreviate
\begin{equation}
	A_{ij} = (\nu_{ij}^\prime-\mu_{ij}^\prime)\delta z_j
		-(\nu_{ij}-\mu_{ij})\delta z_i \ ; 
\end{equation}
\begin{equation}
	B_{ij} = \mu_{ij}^\prime\delta z_j + b_j+a_jt_j
	-\mu_{ij}\delta z_i -b_i-a_it_i
	\ .
\end{equation}
For brevity, we define relative fractions
  $\mu_{ij}$, $\nu_{ij}$, $\mu_{ij}^\prime$, and $\nu_{ij}^\prime$
  expressing how far the overlap occurs along the segments $i$ and $j$.
\begin{eqnarray}
	\mu_{ij}
	\hspace{-3mm}&\equiv&\hspace{-3mm}
	{{T_{ij}-t_i}\over{u_i-t_i}} \ ; 
\nonumber\\
	\nu_{ij}
	\hspace{-3mm}&\equiv&\hspace{-3mm}
	{{U_{ij}-t_i}\over{u_i-t_i}} \ ; 
\nonumber\\
	\mu_{ij}^\prime
	\hspace{-3mm}&\equiv&\hspace{-3mm}
	{{T_{ij}+\tau-t_j}\over{u_j-t_j}} \ ; 
\nonumber\\
	\nu_{ij}^\prime
	\hspace{-3mm}&\equiv&\hspace{-3mm}
	{{U_{ij}+\tau-t_j}\over{u_j-t_j}}
	\ .
\end{eqnarray}

\subsection*{A2. Structure functions for unevenly sampled data}

For continuous data, the order-$n$ structure function is defined as 
\begin{eqnarray}
S_n(\tau) &  \equiv & \left\langle
		\left[{	z(t+\tau)-z(t)	}\right]^n
	\right\rangle \nonumber \\ 
 &	= & 
	\frac{1}{\mathcal T} \int_{\mathcal T} [z(t+\tau)-z(t)]^n\,{\mathrm d}t
	\ .
\end{eqnarray}
For patchy, discretely binned data, we need to integrate contributions from
   each pair of potentially overlapping bins,
   piece by piece.
If bin $i$ overlaps with the offset copy of bin $j$
   then this pair contributes an amount $I_{ij}$
   to the numerator of the structure function,
\begin{eqnarray}
	I_{ij}
	&=& I_{ij}(\tau)
	=\int_{T_{ij}}^{U_{ij}}\left[{
		z^\prime(t+\tau) - z(t)
	}\right]^n\,{\mathrm d}t
\nonumber\\
	&=&
	\int_{T_{ij}}^{U_{ij}}\left[{
		a_j\,t+a_j\,\tau+b_j -a_i\,t -b_i
	}\right]^n\,{\mathrm d}t
	\ ,
\label{eq.general.Iij}
\end{eqnarray}
so that $S_n = S_n(\tau) = W^{-1}\sum_{ij}I_{ij}$.
The normalisation factor $W$ is the total duration
   of temporal overlaps between all pairs of bins.
\begin{equation}
	W\equiv\sum_{i,j}w_{ij}
		\ .
\end{equation}
For $a_i\neq a_j$ (and $A_{ij}\neq0$), 
   the integral (\ref{eq.general.Iij}) can be evaluated directly,
\begin{eqnarray}
	I_{ij} & = & w_{ij}\int_0^1\,y^n\,{\mathrm d}\lambda \nonumber \\ 
	& = & {{w_{ij}}\over{(n+1)A_{ij}}}\left[{
		(A_{ij}+B_{ij})^{n+1} - B_{ij}^{n+1}
	}\right]  \ . 
\label{eq.Iij.integral}
\end{eqnarray}
An alternative, binomial expansion
   proves to be more numerically stable in practice,
   especially where $|A_{ij}|$ is small:
\begin{equation}
	I_{ij} =w_{ij}\sum_{k=0}^n
		{n \choose k}
		{{A_{ij}^k B_{ij}^{n-k}}\over{k+1}}
	\ .
\end{equation}

\subsection*{A3. Determination of the uncertainties}

If each bin $k$ has a measurement $z_k$,
   with uncertainty $\Delta z_k$,
   then we obtain the uncertainty on the overall structure function by quadrature.
Now to propagate the uncertainties in the measurements $z_k$
   to uncertainties in the structure functions $S_n(\tau)$,
   one needs partial derivatives of each piece $I_{ij}$ 
   with respect to each observable $z_k$.
It emerges that
\begin{eqnarray}
	{{\partial I_{ij}}\over{\partial z_k}}
	\hspace{-3mm}&=&\hspace{-3mm}
	w_{ij}\sum_{k=0}^n{n \choose k}
	\biggl[
		{k\over{k+1}}
		A_{ij}^{k-1} B_{ij}^{n-k}
		{{\partial A_{ij}}\over{\partial z_k}}
\nonumber\\&&\hspace{20mm}
		+{{n-k}\over{k+1}}A_{ij}^k B_{ij}^{n-k-1}
		{{\partial B_{ij}}\over{\partial z_k}}
	\biggr]
	\ .
\end{eqnarray}
If we assume a zigzag data patching scheme ($z_{i+1}=z_i+\delta z_i$)
then the revelant partial derivatives of $A_{ij}$ and $B_{ij}$ are
\begin{eqnarray}
	{{\partial A_{ij}}\over{\partial z_k}}
	&=&   \hspace*{2.5mm}
	(\nu_{ij}^\prime-\mu_{ij}^\prime)
	\delta_{_{k,j+1}}	-
	(\nu_{ij}^\prime-\mu_{ij}^\prime)
	\delta_{_{k,j}}
\nonumber\\
   && 	- (\nu_{ij}-\mu_{ij})
	\delta_{_{k,i+1}}
	+
	(\nu_{ij}-\mu_{ij})
	\delta_{_{k,i}} \ ;   \\  
	{{\partial B_{ij}}\over{\partial z_k}}
	& = &  \hspace*{3mm}
	\mu_{ij}^\prime
	\delta_{_{k,j+1}}
	+
	(1-\mu_{ij}^\prime)
	\delta_{_{k,j}} \nonumber \\ 
	& & 
	-
	\mu_{ij}
	\delta_{_{k,i+1}}
	+
	(\mu_{ij}-1)
	\delta_{_{k,i}} \  , 
\end{eqnarray}
where $\delta$ is the Kronecker delta symbol.
Now abbreviate
\begin{eqnarray}
	{\mathcal A}_{ij}
	&\equiv&
	w_{ij}\sum_{k=0}^n{n \choose k}
	{{k}\over{k+1}} A_{ij}^{k-1} B_{ij}^{n-k} \ ; 
\\
	{\mathcal B}_{ij}
	&\equiv&
	w_{ij}\sum_{k=0}^n{n \choose k}
	{{n-k}\over{k+1}} A_{ij}^k B_{ij}^{n-k-1} \ , 
\end{eqnarray}
and
\begin{eqnarray}
C_{1_{ij}}
	&=&
	{\mathcal A}_{ij}(\nu_{ij}^\prime-\mu_{ij}^\prime)
	+{\mathcal B}_{ij}\mu_{ij}^\prime \ ; 
\\
C_{2_{ij}}
	&=&
	-{\mathcal A}_{ij}(\nu_{ij}^\prime-\mu_{ij}^\prime)
	+{\mathcal B}_{ij}(1-\mu_{ij}^\prime) \ ; 
\\
C_{3_{ij}}
	&=&
	-{\mathcal A}_{ij}(\nu_{ij}-\mu_{ij})
	-{\mathcal B}_{ij}\mu_{ij} \ ; 
\\
C_{4_{ij}}
	&=&
	 {\mathcal A}_{ij}(\nu_{ij}-\mu_{ij})
	+{\mathcal B}_{ij}(\mu_{ij}-1) \  . 
\end{eqnarray}
It follows that 
\begin{equation}
	{{\partial I_{ij}}\over{\partial z_k}}
	=
	C_{1_{ij}}\delta_{_{k,j+1}}
	+C_{2_{ij}}\delta_{_{k,j}}
	+C_{3_{ij}}\delta_{_{k,i+1}}
	+C_{4_{ij}}\delta_{_{k,i}} \ . 
\label{eq.dIdz.general}
\end{equation}
By contracting the Kronecker deltas, we obtain 
\begin{eqnarray}
	{{\partial S_n}\over{\partial z_k}}
	\hspace{-3mm}&=&\hspace{-3mm}
	{1\over{W}}
	\sum_{ij}{{\partial I_{ij}}\over{\partial z_k}}
\nonumber \\
	\hspace{-3mm}&=&\hspace{-3mm}
	{1\over{W}}\sum_i\left({
		C_{1_{i,k-1}} +C_{2_{i,k}}+C_{3_{k-1,i}} +C_{4_{k,i}}
		}\right)
	\ .  
\label{eq.dSdz.general}
\end{eqnarray}
The four matrices (${\sf C}_1, {\sf C}_2, {\sf C_3}, {\sf C}_4$)
   only need to be computed once for each choice of $\tau$.
If the uncertainties in the flux measurements are $\Delta z_k$,
  the total uncertainty in the structure function is then given by
\begin{equation}
	\left({\Delta S_n
	}\right)^2
	=
	\sum_k\left({
	{{\partial S_n}\over{\partial z_k}}
	\Delta z_k
	}\right)^2
	\ .
\label{eq.sf.uncertainty}
\end{equation}

\bibliographystyle{mn2e}
\bibliography{jour,dizzy}

\label{lastpage}
\end{document}